\documentclass[apjl]{emulateapj}

\usepackage{natbib}
\usepackage{amsmath,amssymb}
\usepackage[english]{babel}
\usepackage{booktabs}
\usepackage{graphicx}
\usepackage{appendix}

\def\vmax{\ifmmode {v_{\rm max}} \else $v_{\rm max}$\fi}
\def\rmax{\ifmmode {r_{\rm max}} \else $r_{\rm max}$\fi}

\def\vs{\ifmmode {v_s} \else $v_s$\fi}
\def\rs{\ifmmode {r_s} \else $r_s$\fi}
\def\mds{\ifmmode {M_{ds}} \else $M_{ds}$\fi}
\def\nds{\ifmmode {n_{ds}} \else $n_{ds}$\fi}
\def\rods{\ifmmode {\rho_{ds}} \else $\rho_{ds}$\fi}
\def\m300{\ifmmode {M_{300}} \else $M_{300}$\fi}
\def\mm150{\ifmmode {M_{150}} \else $M4_{150}$\fi}
\def\x300{\ifmmode {x_{300}} \else $x_{300}$\fi}

\def\mvir{\ifmmode {M_{vir}} \else $M_{vir}$\fi}
\def\rvir{\ifmmode {r_{vir}} \else $r_{vir}$\fi}
\def\sig{\ifmmode {\sigma} \else $\sigma$\fi}

\def\tgas{\ifmmode {T_{\rm gas}} \else $T_{\rm gas}$\fi}
\def\rgas{\ifmmode {r_{\rm gas}} \else $r_{\rm gas}$\fi}
\def\rogas{\ifmmode {\rho_{\rm gas}} \else $\rho_{\rm gas}$\fi}
\def\mgas{\ifmmode {M_{\rm WM}} \else $M_{\rm WM}$\fi}
\def\mhi{\ifmmode {M_{\rm HI}} \else $M_{\rm HI}$\fi}
\def\nhi{\ifmmode {N_{HI}} \else $N_{HI}$\fi}
\def\rhi{\ifmmode {R_{HI}} \else $R_{HI}$\fi}
\def\rhalf{\ifmmode {r_{1/2}} \else $r_{1/2}$\fi}
\def\thalf{\ifmmode {\theta_{1/2}} \else $\theta_{1/2}$\fi}
\def\npeak{\ifmmode {N_{peak}} \else $N_{peak}$\fi}

\def\cg{\ifmmode {c_g} \else $c_g$\fi}
\def\c6{\ifmmode {c_{g,6}} \else $c_{g,6}$\fi}
\def\phim{\ifmmode {P_{\rm HIM}} \else $P_{\rm HIM}$\fi}

\def\kb{\ifmmode {\it k_{\rm B}} \else $\it k_{\rm B}$\fi}
\def\mp{\ifmmode {m_{p}} \else $m_{p}$\fi}

\def\chisq{\ifmmode {\chi_{mod}^2} \else $\chi_{mod}^2$\fi}

\def\msun{\ifmmode {\rm M_{\odot}} \else $\rm M_{\odot}$\fi}
\def\lsun{\ifmmode {\rm L_{\odot}} \else $\rm L_{\odot}$\fi}

\def\kms{\ifmmode {\rm \:\: km~s^{-1}} \else $\rm km~s^{-1}$\fi}
\def\cmc{\ifmmode {\rm \:\: cm^{-2}} \else $\rm cm^{-2}$\fi}
\def\cmv{\ifmmode {\rm \:\: cm^{-3}} \else $\rm cm^{-3}$\fi}
\def\kel{\ifmmode {\rm \:\: K} \else $\rm K$\fi}
\def\pc{\ifmmode {\rm \:\: pc} \else $\rm pc$\fi}
\def\kpc{\ifmmode {\rm \:\: kpc} \else $\rm kpc$\fi}
\def\amu{\ifmmode {\rm \:\: amu} \else $\rm amu$\fi}

 \shorttitle{UCHVCs as Minihalos}
 \shortauthors{Faerman~et~al.}
  \slugcomment{Accepted to ApJ}

\begin{document}

\title{Ultra-Compact High Velocity Clouds as Minihalos \\ and Dwarf Galaxies}
\author{
Yakov Faerman \altaffilmark{1},  Amiel Sternberg \altaffilmark{1}
and Christopher F. McKee \altaffilmark{2}
}

\altaffiltext{1}
{Raymond and Beverly Sackler School of Physics \& Astronomy,
Tel Aviv University, Ramat Aviv 69978, Israel. e-mail: yakovfae@post.tau.ac.il}
\altaffiltext{2}
{Department of Physics and Department of Astronomy,
University of California at Berkeley, Berkeley CA 94720, USA}


\begin{abstract}
We present dark-matter minihalo models for the Ultra-Compact High Velocity HI Clouds (UCHVCs) recently discovered 
in the 21~cm ALFALFA survey. We assume gravitational confinement of~$10^4$~K HI gas by flat-cored dark-matter subhalos 
within the Local Group. We show that for flat cores, typical (median) tidally-stripped cosmological subhalos at redshift $z=0$
have dark-matter masses of $\sim 10^7$~\msun~within the central 300 pc (independent of total halo mass), 
consistent with the ``Strigari mass scale" observed in low-luminosity dwarf galaxies. Flat-cored subhalos also 
resolve the mass-discrepancy between simulated and observed satellites around the Milky Way. 
For the UCHVCs we calculate the photoionization-limited hydrostatic gas profiles for any distance-dependent total observed HI mass
and predict the associated (projected) HI half-mass radii, assuming the clouds are embedded in distant ($d\gtrsim300$~kpc) and unstripped subhalos. 
For a typical UCHVC (0.9~Jy~\kms) we predict physical HI half-mass radii of 0.18 to 0.35 kpc 
(or angular sizes of 0.6 to 2.1 arcmin) for distances ranging from 300~kpc to 2~Mpc. 
As a consistency check we model the gas-rich dwarf galaxy Leo~T,  
for which there is a well-resolved HI column density profile and a known distance (420 kpc). 
For Leo~T we find that a subhalo with $\m300=8~(\pm 0.2) \times 10^6$~\msun~best fits the observed HI profile.
We derive an upper limit of $\phim \lesssim 150$~\cmv~K for the pressure of any enveloping hot IGM gas at the distance of Leo~T.
Our analysis suggests that some of the UCHVCs may in fact constitute a population of 21-cm-selected 
but optically-faint dwarf galaxies in the Local Group.

\end{abstract}

\section{Introduction}
\label{sec_intro}

Recent sensitive and high-resolution 21~cm observations, carried out as part of the ALFALFA survey \citep{Gio05,Gio07,Haynes11} 
have revealed a population of isolated Ultra-Compact High-Velocity Clouds (UCHVCs) \citep[hereafter G10]{Gio10}, and catalogued by 
\citealt{Adams13} (hereafter A13). The angular diameters are $\lesssim 10$~arcmin, and the UCHVCs are significantly 
smaller than the $\sim 1^{\circ}$-sized CHVCs,  discussed and analyzed by \citet{BB99} and by \citet[hereafter SMW02]{SMW02}. 
The UCHVCs are isolated kinematically and spatially from the well-studied and extended HVC complexes 
(\citealt{Oort70, WW97, Blitz99b, Tripp03, Bruns04, Binney09, Putman12}; see also \citealt{Saul12}).
The A13 UCHVC catalogue contains $\sim 60$ objects found in data covering around 2,800 square degrees.
The radial velocities of the clouds range from -280 to 270~\kms~(GSR). Around 20 of the UCHVCs are only marginally resolved,
with apparent sizes that may be affected by the Arecibo beam, and these compact sources are the focus of our study.  
The observed 21~cm fluxes range from 0.6 to 1.8 Jy~\kms, and the typical central HI column densities averaged over the $4'$ Arecibo beam are 
$5 \times 10^{18}$ to $2 \times 10^{19} \cmc$. The survey does not provide any distance constraints, and the locations of the UCHVCs are unknown. 
The HI masses scale as the square of the distance, and they the range from $1.4 \times 10^5 d_{\rm Mpc}^2$ to $4.2 \times 10^5 d_{\rm Mpc}^2$~\msun, 
where $d_{\rm Mpc}$ is the distance in Mpc. G10 and A13 have suggested that some of the newly discovered UCHVCs might be low-mass 
optically faint dwarf galaxies that are sufficiently gas rich to be detectable 21~cm sources within the ``Local Volume" or Local Group.

In the Local Group, the number of known dwarf galaxies has increased significantly in recent years with sensitive observations carried out as part of the Sloan Digital Sky Survey (SDSS) \citep{Zuk04,Zuk06a,Willman05,Bel06,Bel07,Walsh07,MC08,Bell11,Slater11,MC12}. 
Most of the optically  faint dwarfs are gas-poor systems with old stellar populations, 
are very dark-matter dominated, and are located within the virial radius of the Milky Way (a few 100 kpc). 
Remarkably, the observed stellar velocity dispersions imply a common DM mass scale for  
the dwarf galaxies (Strigari~et~al~2008, hereafter~S08; Walker~2012). 
For $V$-band luminosities, $\it L_{V}$, from $10^3$ to $10^7$~\lsun, S08 find that the DM masses within 300~pc
of the galaxy centers lie within $4\times10^6$ to $3 \times 10^7~\msun$. 
The characteristic Strigari mass is thus $\m300=10^7~\msun$. 

One SDSS dwarf that does contain significant amounts of gas is Leo~T,  discovered by \cite{Irwin07}, 
and detected in 21~cm in the HIPASS survey \citep{Wong06}.  Leo~T is a dark-matter dominated and non-rotating galaxy.
The optically inferred distance to Leo~T is $420 \pm 20 \kpc$ and the radial velocity is -58~\kms~(GSR).
\citet[hereafter RW08]{RW08} presented high-resolution GMRT and WSRT observations of the HI gas distribution. 
The peak HI column density is $\approx 5.5 \times 10^{20} \cmc$, the projected HI half-mass radius is $\approx 0.17$~kpc (1.4'), and the total HI mass is $\approx 3.0 \times 10^5~\msun$. 
(By definition and throughout this paper, the projected HI half-mass radius equals the observed 21~cm half-flux radius.)
As a 21~cm source, Leo~T is similar to the UCHVCs in angular size, but it has a somewhat higher HI column density.
Within the 4' Arecibo beam the average HI column is $10^{20}$~\cmc.

Recently, one of the UCHVCs, HI102145.0+180501, was observed optically by \citet[hereafter G13]{Gio13} and \citet[hereafter R13]{Rhode13} and found to be an ultra-faint dwarf galaxy. G13 dubbed this galaxy Leo~P. Its radial velocity is 177~\kms~(GSR) and the V-band luminosity is $L_V = 1.47 \times 10^5 d_{\rm Mpc}^2~\lsun $.  The optical observations enable a distance determination and place Leo~P at $\sim 1.5-2$~Mpc, outside but possibly bound to the Local Group. The distance constraints are an important advance in estimating characteristic distances for the UCHVCs. 

For Leo~P, combined Arecibo and EVLA observations give a total HI mass of $3.1 \times 10^5 d_{\rm Mpc}^2$~\msun
~and the peak HI column density is $\approx 1.4 \times 10^{20} \cmc$. 
The angular HI half-mass radius is $\thalf \approx 1'$ (or $0.29d_{\rm Mpc}$~kpc), and the total dynamical mass, 
$M_{\rm dyn}$, within $1'$~is $\sim 8 \times 10^6 d_{\rm Mpc}$~\msun. 
G13 find evidence for significant rotation in Leo~P, with ordered motions across the source.
Leo~P is thus an optical dwarf galaxy discovered as a UCHVC via 21~cm observations. So far it is the only UCHVC in which stars have been detected.

Except for rotation, Leo~P appears to be very similar to Leo~T. Both have young stellar populations and significant amounts of gas  
(with HI to stellar mass ratios of $\sim 4$), and both are dark-matter dominated systems, with DM consituting 85\% of the dynamical mass in Leo~T. 
The similarity of Leo~T to the UCHVCs, and the recent identification of Leo~P with HI102145.0+180501 suggests that at least some and perhaps 
many of the unresolved UCHVCs  are gas-rich dwarf galaxies  but optically very faint. If some of the UCHVCs are Local Group galaxies, this extra population
may help to resolve the ``missing satellite problem" \citep{Klypin99,Moore99,Kravtsov04,S07,Madau08,Tollerud08}.

In this paper we present models for Leo~T and the UCHVCs as spherical, non-rotating, hydrostatically supported HI clouds in gravitationally dominant DM minihalos, or DM dominated dwarf galaxies. 
We model the distribution of the warm neutral medium (WNM, $\sim 10^4$~K), 
including truncation of the neutral gas clouds by the present-day photoionizing metagalactic radiation field. 
For this purpose we employ and extend the minihalo modeling methods presented in SMW02. 

For the minihalos we assume flat-core DM density profiles. We show that for flat-core DM halos, the Strigari mass, 
$\m300 \approx 10^7~\msun$, is naturally expected for (tidally stripped) subhalos at redshift zero, 
and only weakly dependent on their total masses (or scale velocities). We also show that adopting flat-core halos resolves the 
``over-abundance problem" of massive dark satellites around the MW found in simulations.

For Leo~T we explicitly fit the observed HI column density profile, and also show that constraints 
set on the DM subhalo properties by just the HI sizes provide consistent results. Thus Leo~T validates our proposed method of using 
the half-mass HI radius as a constraint on the minihalo parameters for the UCHVCs, when higher resolution observations become available. 
We find that for Leo~T, the best-fitting DM halos have $\m300 = 8~(\pm 0.2) \times 10^6$~\msun.

For the UCHVCs we predict the physical and angular HI sizes given the observed 21~cm fluxes and distance dependent  HI masses. 
With future high-resolution observations of the UCHVC sizes our models can be used to constrain the distances to the sources 
or the properties of the confining DM halos. For example, for a typical barely resolved UCHVC with a 21~cm flux equal 
to 0.9~Jy~\kms~(see A13)~embedded in a typical (median) unstripped subhalo we predict physical HI projected half-mass radii 
of 0.18 to 0.35 kpc (or angular sizes of 0.6 to 2.1 arcmin) for distances between 300 kpc and 2 Mpc.

In \S~\ref{sec_conf} we discuss pressure versus gravitational confinement of the UCHVCs, 
and argue that for large distances pressure confinement is unlikely.
In \S~\ref{sec_models_dm} we describe our minihalo potentials and cosmological scaling relations.
In \S~\ref{sec_mass_scale} we show that subhalos with cored profiles naturally produce the Strigari mass scale 
and also resolve the mass discrepancy between simulated and observed dwarf satellite galaxies of the Milky Way.
Then, in \S~\ref{sec_models_gas} we write down analytic formulae for the WNM HI cloud sizes in terms of the halo parameters.
In \S~\ref{sec_analysis} we present our detailed numerical fits for Leo~T and then present our model predictions
for the physical (and angular) HI half-mass radii of the UCHVCs for a wide range of assumed distances, 
for minihalos around the typical cosmological relation. We discuss and summarize in \S~\ref{sec_summary}.

\section{Confinement and Hot Gas}
\label{sec_conf}

How are the UCHVCs confined? Could they be pressure confined? For distances $\sim 1~\rm Mpc$ this appears unlikely if they are compact ($\thalf \approx 1'$), 
as indicated by the high-resolution EVLA observations of Leo~P. For a pressure-confined (uniform-density) spherical cloud, the central peak column density 
is $N_H = 3.3 \: \rhalf \: n_H	$, where $n_H$ is the gas volume density, and \rhalf~is the (projected) HI half-mass radius.
The average column density over the half-flux radius, $\bar{N} = M_{HI}/2\pi m_H r_{1/2}^2$, is lower than the peak column density, 
and we can derive a lower bound for the gas volume density. For $\rhalf = 0.29 d_{\rm Mpc}$~\kpc~and an HI mass of $2.1 \times 10^5 d_{\rm Mpc}^2$
\msun~(for $\theta_{1/2}=1'$ and a typical UCHVC 21~cm flux of 0.9~Jy~\kms), 
the implied thermal pressure is $P_{\rm th}/\kb=n_H T \gtrsim 200~d_{\rm Mpc}^{-1} \cmv$~K for $\sim 10^4$~K warm neutral medium (WNM).
This may be too large for any (hot) confining medium for Local Group distances, 
since the implied baryonic masses (and X-ray emissivities) would be too large.  For example, for a uniform-density medium, 
or assuming that at a distance $d$ the density of the hot gas (hot ionized medium - HIM) is not larger than the mean density within $d$,
the total hot gas mass is 
\begin{equation} \label{eq:hot}
M_{\rm{HIM}} = 5.2 \times 10^{10} d_{\rm Mpc}^3 P_{\rm th}/\kb T_6 ~~~ \msun ~~~ , 
\end{equation}
where $T_6 = T_{\rm HIM}/10^6$~K.
For $P_{\rm th}/\kb \gtrsim 200~d_{\rm Mpc}^{-1} \cmv$~K, and hot gas at a virial-temperature $T_6\approx 2$, we have that 
$M_{\rm HIM} \gtrsim 5 \times 10^{12}$~\msun~for $d\approx 1$~Mpc,  
comparable to the total dynamical mass of the Local Group (\citealt{Li08}) including the Milky Way and Andromeda. 
Pressure confinement might be viable for UCHVCs at much lower distances, or if $\thalf \gg 1'$. 
Gravitational confinement in dark-matter minihalos is a plausible alternative for UCHVCs that are tracing dwarf galaxies.
For gravitational confinement the pressure at the outer boundary provides a limit on the ambient pressure outside the cloud, 
as we find for Leo~T (\S~\ref{analysis_leot}).

\section{Halo Potentials}
\label{sec_models_dm}

In our study we consider gravitationally-confined models, and adopt flat-core \citep{Burk95} profiles for the dark-matter density distributions and gravitational potentials. 
Our choice of the Burkert profile is motivated by observations of constant density cores in low-mass dwarf galaxies
\citep{Gentile04,Gentile07,Donato04,Spano08,Blok08,Oh11,Gilmore07,Walker11,Jardel12a,Walker13}, 
but see also \citet{Jardel12b} for evidence that not all galaxies have flat cores. 
Furthermore, as we show in \S~\ref{sec_mass_scale}, cored profiles naturally reproduce the Strigari mass 
and also resolve the ``mass discrepancy" or predicted overabundance of massive dark 
subhalos around the Milky Way found in simulations (e.g. \citealt{BK11}, hereafter BK11).

For a spherically symmetric DM density distribution, $\rho_{d}(x) = \rods f_{\rho}(x)$, and for the Burkert profile (SMW02)
\begin{equation}
f^{\rm B}_{\rho}(x) = \frac{1}{(1+x)(1+x^2)} ~~~ ,
\end{equation} 
where $x \equiv r/\rs$, $r$ is the radial distance from the halo center, and \rs~is the DM scale radius. 
For the Burkert profile, the scale density \rods~is equal to the central core density at $x=0$. The enclosed DM mass may be written as
\begin{equation} \label{eq:m1}
M_{d}(x) = \mds f_{M}(x) ~~~ ,
\end{equation}
where the scale mass
\begin{equation} \label{eq:m2}
\mds \equiv \frac{4\pi}{3} \rods \rs^3 ~~~.
\end{equation}
The enclosed mass is
\begin{equation} \label{eq:m3}
f^{\rm B}_{M} (\it x) = \rm \frac{3}{2} \left[ \frac{1}{2} ln(1+{\it x}^2) + ln(1+{\it x}) - tan^{-1}({\it x}) \right] \\
\end{equation}
and can be calculated analytically for any radius. 
The halo scale velocity \vs~is defined via the relation
\begin{equation} \label{eq:vsdef}
v_s^2 \equiv \frac{G\mds}{\rs} =\frac{4\pi}{3} G\rho_{ds} \rs^2 ~~~.
\end{equation}
The maximal circular velocity $\vmax = 0.8 \vs$ and $\vmax$ occurs at $\rmax \simeq 3.25 \rs$.

For comparison, the density and enclosed mass for NFW profiles \citep{NFW97} are given by
\begin{gather} \label{eq:m3b}
	\begin{gathered}
		 	f^{\rm NFW}_{\rho}(x) = \frac{1}{x(1+x)^2} \\
			 f^{\rm NFW}_M({\it x}) = 3 \left[ {\rm ln}(1+x) - \frac{x}{1+x}\right]  ~~~ .
	\end{gathered}
\end{gather}
For NFW, \rods~is the density at $r/\rs \simeq 0.47$, and $\vmax$ occurs at $\rmax \simeq 2.16 \rs$ ($\vmax=0.8\vs$, as for Burkert halos). 
The scale mass and velocity are again given by Equations \eqref{eq:m2} and \eqref{eq:vsdef}.

Given a functional form, $f_{\rho}(x)$, for the density profile,
an arbitrary spherical halo may be selected by specifying any two independent parameters, such as \rs~and \rods. 
However, in general, the halo parameters are correlated in a manner that depends on redshift, environment and cosmological model. 
We adopt the LCDM correlation at redshift $z=0$, as found in the \textit{Aquarius} simulation \citep{Springel08} for 
(a) subhalos in which some of the dark-matter has been tidally stripped by encounters with the MW, and 
(b) for more distant unstripped subhalos (not yet interacting).
The {\it Aquarius} simulations assume collisionless dark-matter, and these naturally give rise to NFW profiles. 
However, as recently shown by \citet{Rocha13} in their study of the flat-cored halos that appear for self-interacting dark-matter, 
the cosmological Springel correlations remain unaltered, and we adopt those here for the Burkert profiles.

For tidally stripped subhalos the {\it Aquarius} correlation may be expressed as
\begin{equation} \label{eq:rmax_corr1}
\rmax = 0.026 \times \vmax^{1.39} 10^{-0.18\sig} ~~~ \kpc ~~~ ,
\end{equation}
where \sig~is the number of standard deviations from the median correlation and \vmax~is in~\kms.
Unstripped subhalos are less concentrated, and
\begin{equation} \label{eq:rmax_corr2}
\rmax = 0.037 \times \vmax^{1.39} 10^{-0.18\sig} ~~~ \kpc ~~~ .
\end{equation}

In our discussion of the MW satellites (\S~\ref{sec_mass_scale}) and our derivation of the Strigari mass we assume stripped subhalos with the view that the gas-poor Sloan dwarfs are all nearby (25 to 250 kpc, S08) 
inside the virial radius of the MW ($250^{+60}_{-30}$~kpc, \citealt{Busha11}), and have plausibly interacted with the Galaxy.
We consider Leo~T and the UCHVCs as more distant objects, and for them we assume the cosmological correlation for unstripped subhalos.

For stripped Burkert halos, Equation~\eqref{eq:rmax_corr1} can be reexpressed as
\begin{equation} \label{eq:rsvs}
r_s = 0.14 \times v_{s,6}^{1.39} 10^{-0.18\sig} ~~~ \kpc ~~~ ,
\end{equation}
where $v_{s,6} \equiv \vs/10 \kms$, which implies that
\begin{equation} \label{eq:ndsvs}
\nds = 11.2 \times v_{s,6}^{-0.78} 10^{0.36\sig}  ~~~ \amu \cmv ~~~ ,
\end{equation}
or
\begin{equation}  \label{eq:ndsrs}
\nds = 13.6 \times \left( \frac{\rs}{0.1 \kpc} \right)^{-0.56} 10^{0.26\sig}  ~~~ \amu \cmv ~~~ ,
\end{equation}
where $\nds \equiv \rods/(1~\rm amu)$ and amu is the atomic mass unit 
\footnote{~$1 \amu \cmv = 1.66 \times 10^{-24}~\rm{g} \cmv = 2.54\times 10^{-2}~\msun~\rm{pc}^{-3}$ = $0.93~\rm{GeV/c^2} \cmv$.}.

For unstripped subhalos
\begin{equation} \label{eq:rsvs2}
	\begin{aligned}
		& r_s = 0.21 \times v_{s,6}^{1.39} 10^{-0.18\sig} ~~~ \kpc \\
		& n_{ds} = 5.4 \times v_{s,6}^{-0.78} 10^{0.36\sig} ~~~ \amu \cmv \\
		& n_{ds} = 8.1 \times \left( \frac{\rs}{0.1 \kpc} \right)^{-0.56} 10^{0.26\sig} ~~~ \amu \cmv ~~~ ,
	\end{aligned}
\end{equation}
only slightly offset compared to stripped subhalos. 
Similar expressions may be developed for NFW profiles, but we do not write them down here.

\section{Common Mass Scale}
\label{sec_mass_scale}

We now show that the Strigari DM mass scale of $10^7$~\msun~within 300~pc arises naturally if stripped flat-core Burkert halos are assumed.
We also show that the predicted overabundance of dark massive subhalos around the MW is moderated with this assumption.

\begin{figure}
\begin{tabular}{c}

 \includegraphics[width=0.45\textwidth]{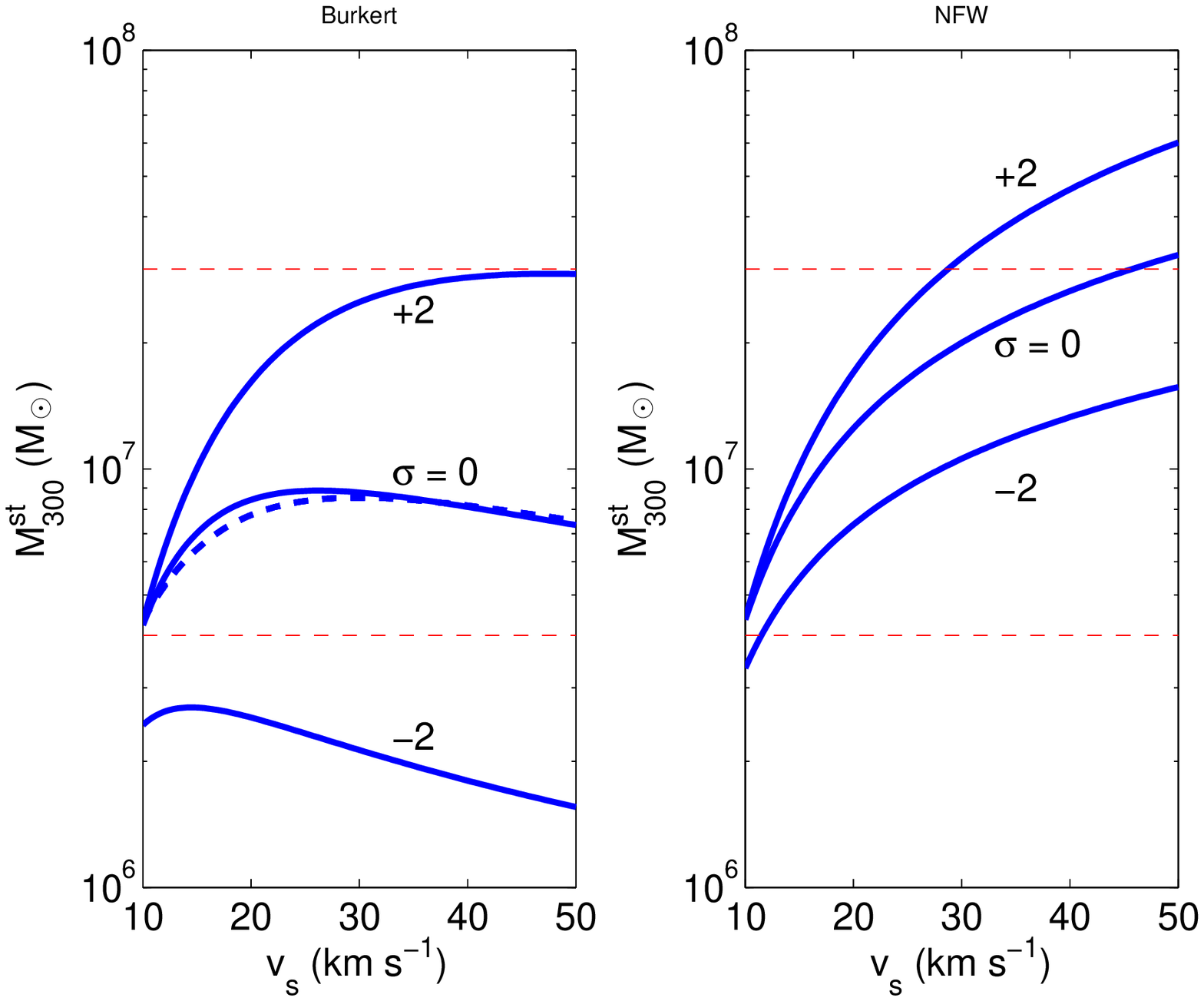} \\
 \includegraphics[width=0.45\textwidth]{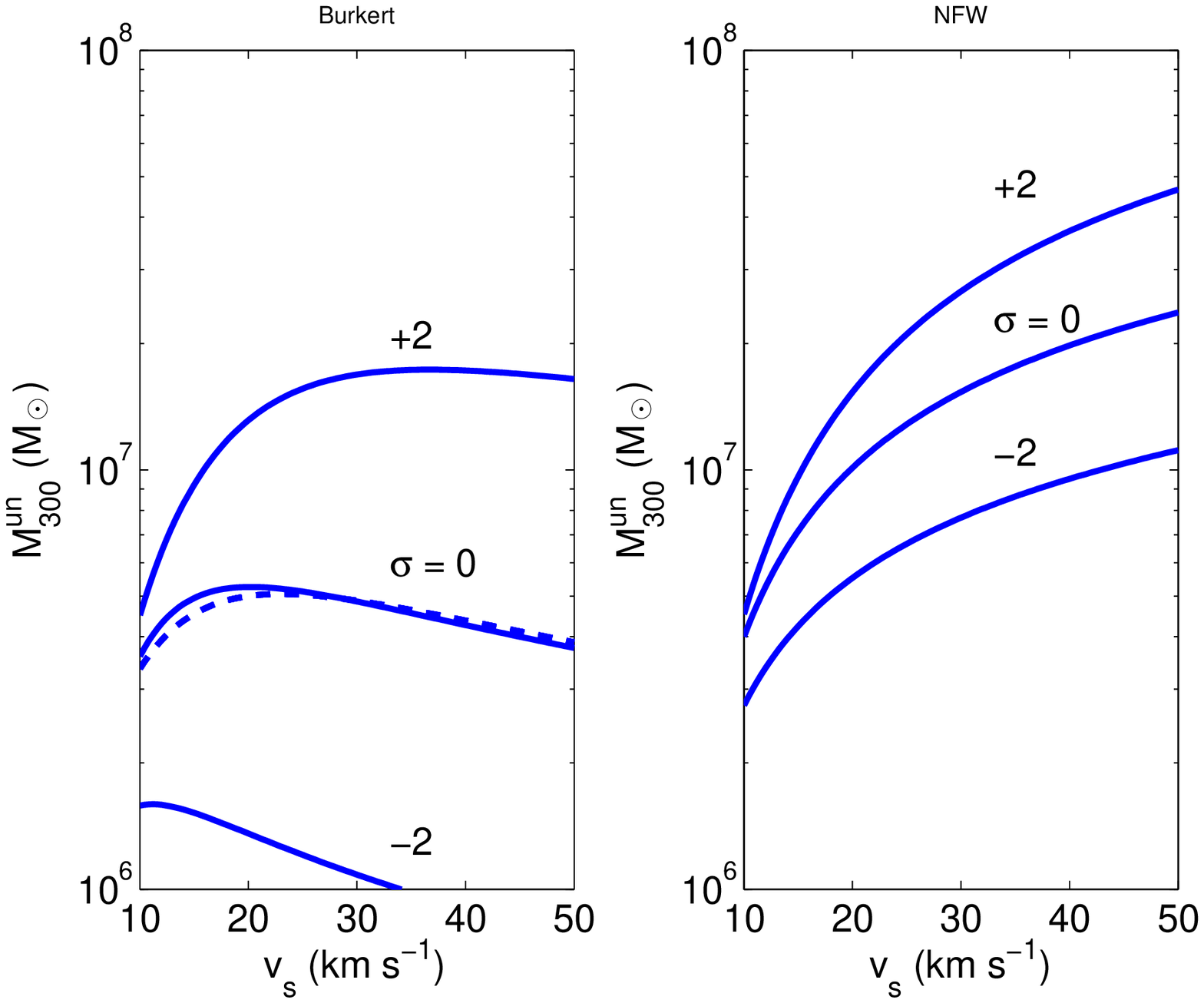} \\
\end{tabular}
\caption{\m300~versus \vs~for stripped (top) and unstripped (bottom) subhalos, with flat-core Burkert (left) and NFW (right) profiles and for \sig=-2, 0, and +2, where \sig~is the number of standard deviations from the median cosmological correlations (Equations [\ref{eq:rmax_corr1}] and [\ref{eq:rmax_corr2}]). The dashed curves in the right panels show the analytic approximations for \m300 for Burkert halos, as given by Equations \eqref{eq:m300b} and \eqref{eq:m300c}. The dashed horizontal lines in the top panels show the observed Strigari range for dwarf satellites of the Milky Way.}
   \label{fig:mass_scale}
 \end{figure}

The total dark matter mass enclosed within a radius $r$ can be written as 
$M_d(r) = (\rs \vs^2 / G) f_{M}(r/\rs)$, with $f_M(x)$ as given by Equations~\eqref{eq:m3} or~\eqref{eq:m3b}. 
Given a cosmological relation between \rs~and \vs, and for a given value of \sig, $M_{r}$ is a function of one halo parameter, 
for any {\it fixed} radius $r$. For Burkert halos this function has a maximum mass, occurring at $x_{300} \equiv 300$~pc/\rs= 0.55.
In contrast, for NFW profiles, the mass within any $r$ varies monotonically. 
In Figure \ref{fig:mass_scale} we plot \m300, the mass within $r = 300$~pc, as a function of the scale velocity, \vs~for Burkert (left panels) 
and NFW halos (right), and for stripped (top panels) and unstripped (bottom) subhalos. 

For stripped subhalos, we find that the maximal \m300~for a given \sig~is 
\begin{equation} \label{eq:m300max1}
\widetilde{M}_{300}^{\rm st} \simeq 8.9 \times 10^6 \times 10^{0.26 \sigma} ~~~ \msun ~~~ ,
\end{equation}
(occurring at $\x300 = 0.55$). For unstripped subhalos, the maximal \m300~is lower, and 
\begin{equation} \label{eq:m300max2} 
\widetilde{M}_{300}^{\rm un} \simeq 5.3 \times 10^6 \times 10^{0.26 \sigma} ~~~ \msun ~~~.
\end{equation}
(The superscripts ``st" and ``un" refer to stripped and unstripped subhalos, respectively.)
The maximum is broad and Equations~\eqref{eq:m300max1} and~\eqref{eq:m300max2} give the ``characteristic" \m300 for any value of \sig.
It is thus evident that for tidally stripped subhalos (Equation~[\ref{eq:m300max1}]), typical (median) Burkert halos are consistent with 
the observed Strigari \m300~range for MW dwarfs with the characteristic mass, $\widetilde{M}_{300}^{\rm st}$, ranging from 
$\sim 3 \times 10^6$ to $\sim 3 \times 10^7~\msun$ for \sig~between -2 and 2. For \sig=0, $\widetilde{M}_{300}^{\rm st} = 8.9 \times 10^6~\msun$, 
and this is essentially the Strigari mass, of $10^7~\msun$. We refer to these $\sig=0$ halos as ``Strigari halos".
For unstripped subhalos, the characteristic mass is lower by a factor of 0.6.
We predict that more distant low-mass dwarfs will have on-average lower characteristic masses 
within their central 300~pc, with $\m300 \approx 5 \times 10^6$~\msun.

As shown in Figure~\ref{fig:mass_scale} (see also \citealt{Maccio09}), NFW subhalos are more massive than Burkert subhalos for any \sig. 
For stripped NFW subhalos (top panel) and for $\sig=0$, \m300~increases by a factor of 7, 
from $4.4 \times 10^6$ to $3.2 \times 10^7$~\msun, for \vs~from 10 to 50~\kms. 
However, for stripped Burkert profiles \m300~varies by less than a factor of 2 across this range relative to 
the characteristic (maximal) mass of $8.9 \times 10^6$~\msun. From 15 to 50~\kms~the variation is less than a factor of 1.3.
For median (stripped) NFW halos the Strigari mass of $10^7$~\msun~occurs for a narrow range of \vs~near 20~\kms~(or $\vmax \sim 16$~\kms).  
For NFW, this has motivated the idea that the common mass scale observed by S08 represents a ``threshold"
with star-formation being significantly inhibited in subhalos with $\vs \lesssim 20$~\kms~and all of the low-luminosity dwarfs
having been formed in subhalos near this threshold \citep{Maccio09, Li09, Okamoto09, Stringer10, Rashkov12}.  
Alternatively, if subhalo cores are flat the observed common mass scale does not necessarily reflect a threshold, 
but could simply be due to the very weak dependence of \m300~on \vs~(or subhalo mass), as shown in Figure~\ref{fig:mass_scale} (left panels).  
For flat-cored DM profiles the scale velocities of the low-luminosity dwarfs 
could span a wide range of \vs~(for $\vs \gtrsim 15$~\kms), all giving the same \m300.
We would then predict that as more faint galaxies are discovered (within the MW virial radius) \m300~will remain invariant.

\begin{figure}
  \includegraphics[width=0.45\textwidth]{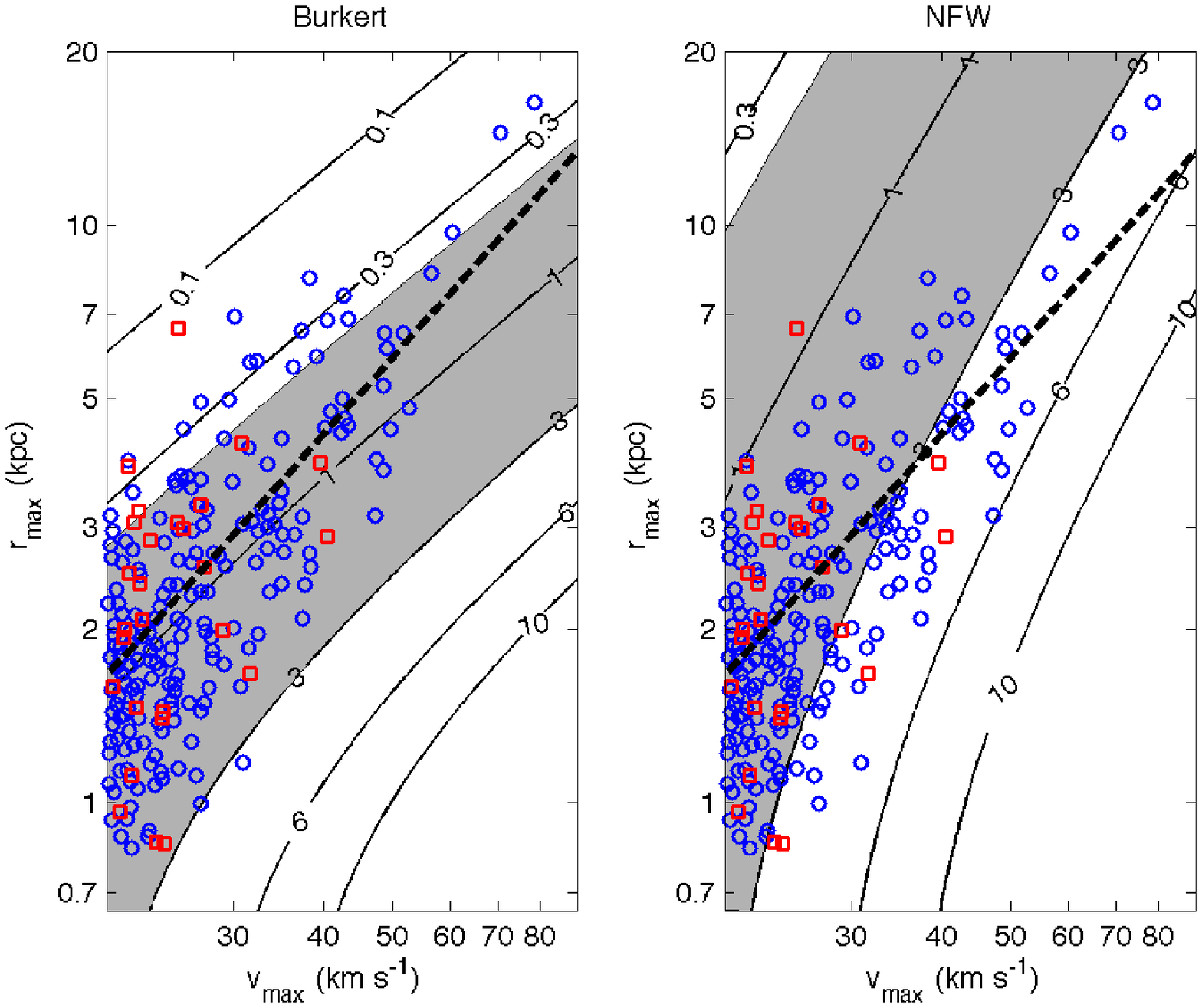}
\caption{Contours of $M_{300}^{\rm st}$~(in units of $10^7~\msun$) for flat-core Burkert halos (left) and NFW halos (right), for \rmax~versus \vmax.
The dashed line is the cosmological halo-parameter-correlation for median stripped subhalos as given by Equation~\eqref{eq:rmax_corr1}.
Blue circles are the \rmax,~\vmax~values for subhalos as extracted by BK11 from the {\it Aquarius} \citep{Springel08} simulation.
The red squares are halos from {\it Via Lactea II} \citep{Diemand07}. 
The shaded regions show the observed range of \m300~for the MW dwarfs, $0.4-3 \times 10^7$~\msun, as found by S08. }
   \label{fig:TBTF}
 \end{figure}

As a further illustration for the difference between Burkert and NFW profiles, 
in Figure~\ref{fig:TBTF} we plot contours for \m300 (in units of $10^7$~\msun) in the \rmax~versus \vmax~parameter plane
for Burkert and NFW profiles (left- and right-hand panels respectively).  For a given \rmax~and \vmax~it is again apparent that NFW halos are
significantly more massive than Burkert halos.  The solid dashed line is the median {\it Aquarius} correlation between 
\rmax~and \vmax~for stripped subhalos as given by Equation~\eqref{eq:rmax_corr1}. For Burkert halos this line is 
very close to the $\m300=10^7~\msun$~contour. However, median NFW halos are much more massive.

The points shown in Figure \ref{fig:TBTF} are the \rmax~and \vmax~values for the accreted subhalos in 
the {\it Aquarius} (blue) and {\it Via Lactea II} (red) simulations. BK11 plotted these data for NFW halos (their Fig.~2) 
as we reproduce in the right-hand panel of our Figure \ref{fig:TBTF}.
For NFW profiles many of the simulated  subhalos are significantly more massive than the observed dwarf galaxies as specified by the Strigari mass range. 
BK11 then postulated that the Local Group may contain an additional population of starless (and gasless) 
subhalos that are more massive than the observed low-mass dwarf galaxies. For simple ``abundance-matching" such massive subhalos are unexpected.
However, they also noted that alternate dark-matter profiles would mitigate this conjecture.  
Our Figure~\ref{fig:TBTF} (left panel) shows that for flat-core Burkert profiles the mass discrepancy disappears.
For flat-core profiles all of the simulated halos are within the observed Strigari mass range, and an extra population of dark massive subhalos is not necessary.  
Flat cores could be produced by self-interacting dark-matter (\citealt{Vogel12, Rocha13}) and as noted by these authors, 
the resulting lowered densities in the cores reduce the mass-discrepancy. We have shown that the characteristic mass is obtained naturally 
and the mass discrepancy is resolved if a flat-core profile is adopted independent of its origin,
so long as the cosmological relation between \rmax~and \vmax~is retained.
However, the number of predicted subhalos at any given \vs~is still larger than observed.  
We argue that some of these missing galaxies may be observable as the UCHVCs, as suggested by G10 and A13.

Our use of the Burkert profile is thus motivated by (a) observational evidence for flat DM cores in dwarf galaxies, 
(b) the fact that such profiles naturally provide the Strigari mass (Equation~[\ref{eq:m300max1}]) for close to  median halos, 
and (c) resolution of the mass discrepancy between simulated subhalos and  observed dwarf galaxy masses. 
We adopt the flat-cored profiles in our modeling of the gas distributions in Leo~T and the UCHVCs, 
as described in \S~\ref{sec_models_gas} and \S~\ref{sec_analysis}.

To conclude this section, it is useful to develop some simple expressions for \m300, for Burkert profiles.
For unstripped subhalos, Equation~\eqref{eq:rsvs} shows that for $\sig \lesssim 3$ and $\vs \gtrsim 20 \kms$, the scale radius 
is $\rs \gtrsim 0.1 \kpc$, so that for realistic halos $\x300 \lesssim 3$. For $x \lesssim 3$,  the enclosed mass function 
as given by Equation~\eqref{eq:m3} is well approximated by the expression
\begin{equation}
f_{M} \approx \frac{x^3}{1+1.87x^{1.60}} ~~~ ,
\end{equation}
which is accurate to within 15$\%$ up to $x=3$. For median (\sig = 0) subhalos it
follows from Equations~\eqref{eq:m1} and \eqref{eq:rsvs} that
\begin{equation} \label{eq:m300a}
M_{300}^{\rm st} \simeq 2.77\times10^7 \frac{\nds/10\amu}{1+4.50(\nds/10\amu)^{2.85}} ~~~ \msun ~~~,
\end{equation}
or in terms of \vs
\begin{equation} \label{eq:m300b}
M_{300}^{\rm st} \simeq 1.80 \times10^7 \frac{(\vs/20 \kms)^{-0.78}}{1+1.32(\vs/20\kms)^{-2.22}} ~~~ \msun ~~~.
\end{equation}
For unstripped subhalos, the approximations are
\begin{equation} \label{eq:m300c}
	\begin{aligned}
	& M_{300}^{\rm un} \simeq 2.77\times10^7 \frac{\nds/10\amu}{1+20.1(\nds/10\amu)^{2.85}}  ~~~ \msun ~~~ \\
	& M_{300}^{\rm un} \simeq 0.87 \times10^7 \frac{(\vs/20 \kms)^{-0.78}}{1+0.74(\vs/20\kms)^{-2.22}} ~~~ \msun ~~~.
	\end{aligned}
\end{equation}
In Figure \ref{fig:mass_scale} we plot these expressions for \m300(\vs)~as the dashed curves in the left-hand panels.

\section{HI Gas}
\label{sec_models_gas}

We model the HI gas distributions in Leo~T and the UCHVCs as thermally supported, non-rotating, hydrostatic gas spheres that are gravitationally 
confined by DM minihalos. 
Table \ref{tab:obs_par} summarizes some of the basic observational properties for Leo~T (RW08) and a representative compact object from the UCHVC catalog 
(see A13), including the (distance-dependent) total HI masses, the half-mass radii, the peak HI column densities and the 21~cm line widths.  
In \S~\ref{sec_gas_an} we write down analytic formulae that can be used to constrain the halo parameters, including \nds~and \m300, 
given the observed HI properties. In \S~\ref{sec_analysis} we present detailed numerical fits for the full HI column density profile and half-mass HI radius in Leo~T,
followed by predicted distance-versus-size computations for the UCHVCs.

\begin{table}
	\caption{Observed parameters for Leo~T and UCHVCs}
	\label{tab:obs_par}
		\begin{tabular}{| l || c c|}
			\toprule
			 & Leo~T 
			 \footnote{Observations of Leo~T are reported in RW08.}
			 \footnote{For Leo~T we seperated the WNM component from the total profile, so we state: total value (WNM value).}
			 & UCHVCs \\
			\midrule
			d [kpc] & $420 \pm 20$ & --- \\
			$M_{HI} [M_{\odot}] $ & $3.0 (1.8) \times 10^5 $ & $2.1 \times 10^5 d_{\rm Mpc}^2$\\
			$\theta_{1/2}$ [arcmin] & 1.4 & $< 2$\\
			$ r_{1/2} [{\rm kpc}] $ & 0.17 & $< 0.58d_{\rm Mpc}$ \\
			$N_{peak} [{\rm cm^{-2}}]$ & $5.5 \: (2.1) \times 10^{20}$ & $\bar{N} = 10^{19}$\\
			$\rm \sigma_{v}$ [\kms] & 6.9 & $ 9.8 \pm 1.3 $\\
			W50 or FWHM [\kms] & 16.6 & $23 \pm 3$ \\
			$\rm L_V ~[L_{\odot}] $ & $6 \times 10^4$ & --- \\
			$v_{\rm GSR}$~[\kms] & -58 & -230 to +230 \\
			\bottomrule
		\end{tabular}
\end{table}

For these purposes we employ and extend the modeling methods described in SMW02.
The HI gas spheres are assumed to be WNM with temperatures $T \sim 10^4 \kel$, and surrounded by photoionized warm ionized medium (WIM) 
shielding envelopes with comparable temperature.  The assumed temperature accounts for any turbulent motions if present.
The gas sound speed, $c_g$, is an important parameter as it is directly related to the observed 21~cm line width. For isothermal gas, \cg = FWHM/$2\sqrt{2\ln{2}}$. 
For $T = 10^4 \kel$, $c_g$ equals $8.19$ and $11.6 \kms$ for the WNM and WIM. We assume rotational support is negligible.

Photoionization and radiative heating are by the externally incident present-day ($z=0$) UV/X-ray metagalactic field. 
We use the SMW02 representation (their Appendix B) for the metagalactic radiation, which is very similar to the updated \citet{HM12} 
spectrum for $z$=0. Internal stellar radiation is not included. 
The radiative transfer and the coupled hydrostatic and ionization structures are computed self consistently, 
and conditions are determined for the possible existence of multiphased warm/cold (WNM/CNM) cores, 
and for complete conversion to a cold neutral medium (CNM, $T \lesssim 200 \kel$) at the cloud centers. For a given halo potential, the total mass, 
\mgas,~of warm gas (WNM plus WIM) is the primary free parameter, and determines the gas fractions 
in the concentric WIM, WNM and WNM/CNM zones. For sufficiently small \mgas~the cloud is fully ionized. 
As \mgas~is increased, an inner HI WNM sphere appears, followed by the formation of a multiphased WNM/CNM core.

An additional model parameter is the external pressure, possibly provided by an unbound (hot) medium (HIM). 
However, as discussed in \S~\ref{sec_conf} it is unlikely that the HI gas in the UCHVCs is pressure confined if they are at large distances. 
In our study we assume that any external bounding pressure is negligible, and that gravitational confinement dominates. 
As we discuss in \S~\ref{sec_analysis}, this requires that the pressure $\phim/\kb \lesssim 150 \cmv$~K.

\subsection{Analytic Results for HI in the Small-$x$ Limit (Unstripped Subhalos)}
\label{sec_gas_an}

For sufficiently deep gravitational potential-wells the gas is confined to the inner parts of the halos, where $x \equiv r/\rs$ is small. 
In the ``small-$x$ limit", convenient analytic expressions are available (SMW02) for the radial gas density distributions, 
and for the associated gas scale heights. 

\begin{figure}
 \includegraphics[width=0.45\textwidth]{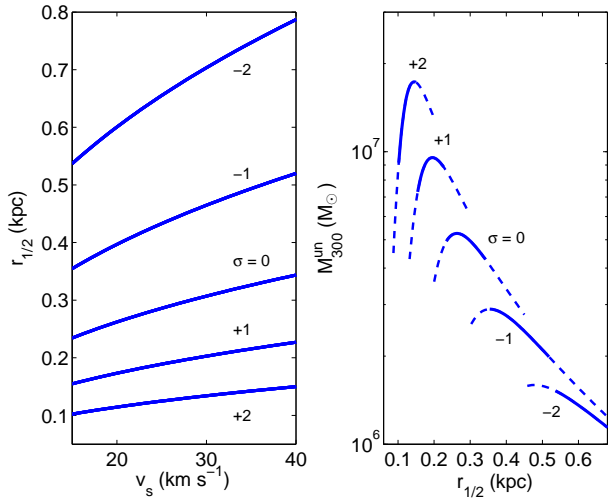}
\caption{
Left: The projected half-mass radius, \rhalf, versus halo scale velocity, \vs, as given by the small-$x$ expression in Equation~\eqref{eq:rgas2}, 
for unstripped subhalos with \sig~between -2 and +2, and sound speed \cg=8.2~\kms. 
Right: $M_{300}^{\rm un}$~versus \rhalf, for \sig~between -2 and +2 and \cg=8.2~\kms. The solid portions of the curves are for \vs~from 15 to 40~\kms.}
   \label{fig:gas}
 \end{figure}

For Burkert potentials, and for fixed gas velocity dispersion, the gas density profile in the small-$x$ limit is a Gaussian
\begin{equation} \label{eq:rgas1}
\rogas(r) = \rogas(0)~{\rm exp} \left[ -\left( \frac{r}{\rgas} \right)^2\right]
\end{equation}
where the gas scale height is $r_{\rm gas} \equiv \sqrt{2}\cg \rs / \vs$.
For a Gaussian gas distribution the projected half-mass radius $\rhalf = \sqrt{\rm ln(2)} \times \rgas$ so that using Equation \eqref{eq:vsdef} 
and with Equation~\eqref{eq:rsvs2} relating \rs~and \vs~it follows that for unstripped Burkert subhalos
\begin{equation}  \label{eq:rgas2}
	\begin{aligned}
   	 	r_{1/2} & = 0.18 \c6 \left(\frac{\nds}{10\amu}\right)^{-1/2} \kpc ~~~ , \\
  			   	 	& = 0.32 \c6 \left(\frac{\vs}{20\kms}\right)^{0.39} 10^{-0.18 \sig} \kpc ~~~ , \\
	 \end{aligned}
\end{equation}
where $\c6 = \cg/10 \kms$. In Figure~\ref{fig:gas} (left) we plot \rhalf~versus \vs~as given by Equation~\eqref{eq:rgas2} for 
underconcentrated, median, and overconcentrated halos, from \sig=-2 to +2, for a sound speed of 8.19~\kms~(\tgas=$10^4$~K).

For a fully neutral gas cloud with fixed velocity dispersion, and not limited by photoionization of the outer envelope, 
the half-mass radius diverges beyond the small-$x$ limit as \vs/\cg~becomes small. 
The small-$x$ expression for the gas scale height $r_{\rm gas}=\sqrt{2}\rs/\vs$~ is accurate
to within 35\% for $\vs/\cg>2$, and 10\% for $\vs/\cg>3$. In \S~\ref{analysis_uchvc} we present computations 
of the projected half-mass radii including external photoionization that truncates the neutral cores depending on the total gas mass.
(In the absence of photoionization \rhalf~diverges more rapidly than \rgas~with decreasing \vs/\cg.) 
For the full range of $\vs/\cg$ (from 2.4 to 4.8) and HI masses that we consider in \S~\ref{analysis_uchvc}
 we find that Equation~\eqref{eq:rgas2} is accurate to better than a factor-two.

In Figure~\ref{fig:gas} (right) we plot \m300 versus \rhalf, for \rhalf~as given by Equation~\eqref{eq:rgas2} for \cg =8.19~\kms. The solid portions of the curves correspond to \vs~from 15 to 40~\kms. For median (\sig=0) halos it follows from Equation~\eqref{eq:m300a} that
\begin{equation} \label{eq:m300d}
M_{300}^{\rm un} \simeq 2.18 \times 10^7 \frac{[\c6 / (r_{1/2}/0.2 \kpc)]^2}{1+10.0[\c6 / (r_{1/2}/0.2 \kpc)]^{5.70}} \\
\end{equation}
The maximum, $ \widetilde{M}_{300} = 5.3 \times 10^6~\msun$, occurs when $\rhalf \simeq 0.27$~kpc.
For $\rhalf > 0.27 \kpc$, a radius of 300~pc is within the constant density core ($x_{300}<1$), 
so $\m300 \propto r_{1/2}^{-2}$ (Equation [\ref{eq:rgas2}]). For $\rhalf < 0.27 \kpc$, $\x300 > 1$~and although 
\nds~increases with decreasing \vs, the mean DM density within 300~pc decreases and therefore so does \m300.

Writing $\rhalf = d \times \theta_{1/2}$, where  $\theta_{1/2}$ is the angular projected HI half-mass radius, 
we can express the distance as a function of the angular radius, the sound speed, and halo parameters
\begin{equation} \label{eq:d1}
	\begin{aligned} 
	d_{\rm Mpc}  & = 0.61 \c6 \left(\frac{\theta_{1/2}}{1'} \right)^{-1} \left(\frac{\nds}{10 \amu \cmv}\right)^{-1/2}  \\
							&= 1.1 \c6 \left(\frac{\theta_{1/2}}{1'} \right)^{-1} \left(\frac{\vs}{20 \kms}\right)^{0.39} 10^{-0.18\sig}
	\end{aligned}
\end{equation}
where $d_{\rm Mpc}$ is the distance in Mpc.

As we show in \S\ref{analysis_uchvc}, the dependence of the distance on \vs~is actually reduced by departures from the small-$x$ limit.  
A dependence on the HI gas mass also enters because the gas mass affects the positions of the photoionization cut-offs, not included in our 
small-$x$ formulae above.  As we will show, for a given 21~cm flux and angular size the implied distance depends almost entirely on the halo \sig.
However, Equation \eqref{eq:d1} may be used for some simple analytical estimates, as follows.

We expect that for thermally stable WNM, the sound-speed is restricted to a narrow range (8.2~\kms~for $10^4$~K). 
For the UCHVCs, our Equation~\eqref{eq:d1} provides distance estimates to the source given an observed angular size, \thalf, for an assumed \sig~and \vs.  
For example, for a median (unstripped) subhalo and assuming \thalf = 1' (and \cg=8.2~\kms), the distance is 0.9 to 1.2 Mpc, for \vs~between 20 and 40 \kms.
Turning this around, for a known distance, we can constrain the halo properties. For example, for Leo~T, with $\theta_{1/2} = 1.4'$ and 
\cg=7~\kms~(see Table \ref{tab:obs_par}), the observed distance of 420~kpc implies that $\sig \approx +1$ for which $\widetilde{M}_{300}^{\rm un} \sim 10^7$~\msun.  
This analytic result is in harmony with our more detailed numerical fitting for the HI profile in Leo~T (\S~\ref{analysis_leot}).

What about the larger-sized CHVCs originally considered by SMW02 ? 
The CHVC diameters are $\gtrsim 24'$ (\citealt{BBC01,Putman02}, A13), and for these we estimate HI half mass radii of $6'$~to~$8'$.
In the minihalo interpretation, Equation \eqref{eq:d1} implies distances of $\leq 200$~kpc (for median halos) for the CHVCs, as concluded by SMW02.
However, the recently observed gas-poor Sloan dwarfs suggest that within the MW virial radius minihalos would be less likely to retain their gas.
Some CHVCs could be embedded in very underconcentrated halos, with $\sig<-2$
(leading to a larger physical size and thus larger distance), but more likely they are nearby pressure-confined clouds. 
This is supported by the large H$\alpha$ surface brightnesses found by \citet{Tufte02} and \citet{Putman03} 
for some of the CHVCs, indicating distances $<40$~kpc for this population.
However, the smaller UCHVCs, especially the most compact ones could be much more distant objects.

\section{Numerical Models}
\label{sec_analysis}

We now present numerical model computations for the HI gas distributions, gas masses, and half-mass radii for Leo~T and for the UCHVCs. 
In these calculations we do not necessarily restrict ourselves to the small-$x$ limit. 
We also include the effects of external photoionization in limiting the cloud sizes and masses of the neutral HI cores. 
This introduces a dependence of the cloud size on the gas mass that does not appear in our small-$x$ formulae.
For Leo~T and the UCHVCs we use the cosmological relation for unstripped DM subhalos, as given by Equation \eqref{eq:rmax_corr2}, 
since gas-rich systems are expected to reside outside the virial radius of the Milky Way.
We assume isothermal conditions for the WNM and WIM. For Leo~T we assume $6000$~K, as observed, and for the UCHVCs we assume $T=10^4$~K gas. 
We assume that any external bounding pressure, \phim, is very low compared to the resulting central pressures. To ensure this we set $\phim/\kb=10$~\cmv~K. 
With this assumption the external pressure is always negligible for the range of halo parameters that we consider, and the clouds are gravitationally confined.

In \S~\ref{prof_fit}  we construct detailed fits for the observed HI profile in Leo~T.
We find that the \m300~that produces the best fit is $8 \times 10^6~\msun$, 
slightly larger than the typical median characteristic mass expected for unstripped subhalos (Equation [\ref{eq:m300max2}]). Then, in \S~\ref{rhalf_fit} we verify that the projected HI half-mass radii may be used to set constraints on the halo parameters, especially \m300. We constrain the halo scale velocity in \S~\ref{vs_fit}.

In \S~\ref{analysis_uchvc} we model the UCHVCs as minihalo clouds within $\sig=\pm 1$ of typical median halos. We extract the calculated half-mass radii and plot \rhalf~and \thalf~as functions of the assumed distance $d$ to the source. These predictions can be used for interpreting future high-resolution observations of the UCHVCs as minihalos.
An assumed \sig~for the DM halo sets the physical scale size of the HI gas distribution, enabling a distance estimate given an observed angular size. Alternatively, if the distances to these objects are measured by some other method, the gas distributions can be used to constrain the DM halo parameters, as we do for Leo T.

\subsection{Leo~T}
\label{analysis_leot}

RW08 presented GMRT and WSRT mapping observations of the HI in Leo~T.  They detected cold ($\sim 500$~K CNM) and warm 
($\sim 6000$~K WNM) components. The CNM is concentrated in the inner regions within the central $r \sim 1$~arcmin (0.12~kpc).
The WNM extends out to $r \sim 3$ arcmin ($\sim 0.35$~kpc), to the WSRT sensitivity limit of $2 \times 10^{19} \cmc$.  

We are interested in modeling the distribution of the WNM in Leo~T.  
For this purpose we constructed a radially averaged CNM density profile using the GMRT CNM map
presented by RW08 (their Fig.~1). We fit the CNM profile with a second-degree polynomial, 
and subtracted it from the radially averaged total HI column density profile (Ryan-Weber, priv.~comm.). 
The resulting WNM profile for Leo~T is displayed in Figure~\ref{fig:leot_prof} as the (red) solid curve, including error bars (see below).
The total WNM gas mass is $2.8 \times 10^5~\msun~(\pm 10\%)$.

\subsubsection{Fitting the Full HI Profile}
\label{prof_fit}

We fit the observed HI profile in Leo~T as follows. 
First, we can define a DM halo potential by specifying the scale velocity, \vs, and the deviation from the median halo, \sig. 
Equivalently, since \m300~is a known function of \vs~and \sig, we can specify the halo potential in terms of \vs~and \m300.
We then add gas until the total WNM mass is equal to the observed value of $2.8 \times 10^5~\msun$.
We set the WNM gas temperature equal to 6000~K (or \cg=7~\kms), as found by RW08. The resulting hydrostatic cloud structure then yields 
the projected HI column density profile, $N_{\rm HI}( r)$,  where $r$ is the offset radius from the cloud center. 

To compare to the observed profile we define the goodness of fit measure,
\begin{equation}
	\begin{aligned}
		\chi^2 &= \sum_i{\left[\frac{\log(N_{HI,{\rm mod}}(r_i))-\log(N_{HI,{\rm obs}}(r_i))}{\sigma_i} \right]^2} \\
		\sigma_i &= \frac{1}{\ln{10}}\frac{\Delta N_{HI,{\rm obs}}(r_i)}{N_{HI,{\rm obs}}(r_i)} \\
	\end{aligned}
\end{equation}
where $N_{HI,{\rm mod}}(r_i)$, $N_{HI,{\rm obs}}(r_i)$ are the modeled and observed column densities at $r_i$, respectively,
and $\Delta N_{HI,{\rm obs}}(r_i)$ is the error in the observed column density. 
We fit the observed profile, down to the sensitivity level of $\nhi=2\times10^{19} \cmc$. 
In the outer part of the profile we adopt the errors reported in RW08, with $3\sig = 2 \times 10^{19} \cmc$.  
In the inner part ($r<0.12$~kpc) we estimate errors larger by 50\%, due to uncertainties in CNM subtraction.
We vary the minihalo parameters, and the best fitting models are those for which the value of $\chi^2$ is minimized.

\begin{figure}
  \includegraphics[width=0.45\textwidth]{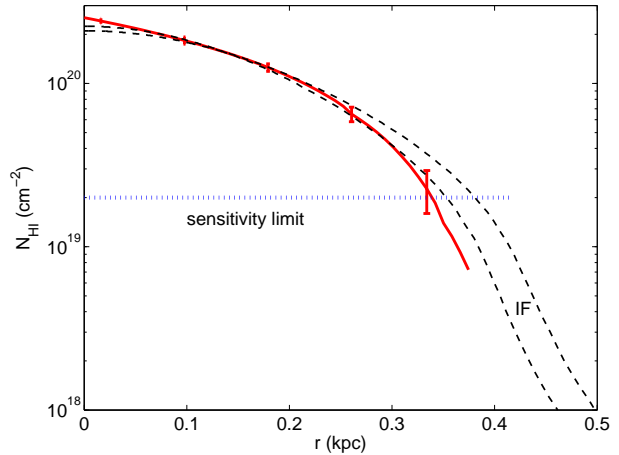}
\caption{The Leo~T HI column density profile. Solid (red) curve is the observed WNM profile as extracted from the RW08 21~cm maps (see text), including the observational error bars. The dashed curves show the best-fitting profiles for $\m300 \sim 8 \times 10^6~\msun$~and \vs~of 18 and 40~\kms, as determined by our $\chi^2$-fitting procedure. The dotted  horizontal line shows the 3\sig~detection sensitivity limit. The ionization fronts, which we define as the radii where the HI (to total H) fraction decreases to 0.5, are at $r \sim 0.42$~kpc.}
   \label{fig:leot_prof}
 \end{figure}

We considered \vs~from 15 to 40~\kms, and \m300 from $2\times 10^6$ to $3\times 10^7~\msun$ (or \sig~from $\sim -2$ to 3 for unstripped subhalos).
As expected, the resulting HI profile is not very sensitive to \vs, but does set strong constraints on \m300~and \sig. The best fits are for \m300 = $8.0~(\pm 0.2) \times 10^6~\msun$ and $\vs=30~(\pm 5)$~\kms. Adopting the cosmological correlation for unstripped subhalos (see Equations [\ref{eq:rmax_corr2}] and [\ref{eq:m300max2}]), this results in a slightly overconcentrated halo, with $\sig \approx 0.75$. The 3\sig~errors on the fitted parameters are $\Delta \m300 \approx 0.5 \times 10^6~\msun$, and $\Delta \vs \approx 12~\kms$. Small variations on the \nhi~error estimates do not change the results significantly. The results of our profile fitting are shown in Figure~\ref{fig:leot_prof}.  
We display the model profiles including the predicted ionization fronts (IF) at $r\approx 0.42$~kpc. Again, the observed WNM profile is the (red) solid curve, with the above mentioned errors.
The dashed curves indicate the range of resulting profiles within $\pm \Delta \vs$ from the best fit. Remarkably, our best fits match the entire observed profile down to the detection sensitivity limit of $2 \times 10^{19} \cmc$. For these fits, the central column density of $2.4 \times 10^{20} \cmc$ is reproduced, and the computed half-mass radius is 0.17~kpc, as extracted from the observed profile. For higher (lower) values of \m300, the central columns are increased (decreased), and the half-mass radii are smaller (larger). For our best fitting models the core densities, \nds, range from 5 to 12 \amu~\cmv~(or 0.12 to 0.30~$\rm\msun~pc^{-3}$). Our best fit models predict the formation of CNM in the centers of the minihalos, consistent with the CNM observed by RW08.

Leo~T can also be modeled using NFW halos, although the best fits are not as good.
For the best fit NFW halo $\m300 \sim 6.5 \times 10^6$~\msun~and $\vs \sim 30$~\kms.
However, and more importantly, the best fitting NFW halos are $\sim -2.5 \sig$, and as such are less likely.

Our result, $\m300=0.8 \times 10^7$~\msun,~is consistent with the measurement by RW08, 
who estimate the total mass within 300~pc to be $>3.3 \times 10^6$~\msun~using the 21~cm line width. 
For Leo~T, S08 found $1.30^{+0.88}_{-0.42} \times 10^7$~\msun~within 300~pc, close to our computed \m300. For our best fits, $\mm150 = 0.12 \times 10^7$~\msun, a factor of 8 smaller than \m300, as expected for constant DM densities within the scale radius, whereas \citet{Wolf10} report $\mm150=0.74^{+0.48}_{-0.29} \times 10^7$\msun~for Leo~T.
However, the mass ratio $\m300/\mm150 \sim 1.75$, as implied by the \citet{Wolf10} and S08 observations, is difficult to reconcile with either Burkert or NFW halos, and may be affected by measurement uncertainties. 

In our fits we have set the external bounding pressure to a low value of \phim/\kb=10~\cmv~K.
At the 0.35~kpc offset corresponding to the detection limit, the computed thermal pressure is $\sim 150 \cmv$~K, and this is then the maximal allowed bounding pressure consistent with our models. For higher bounding pressures the HI gas distribution in Leo~T would be compressed below the observed size. Thus, \phim/\kb=150~\cmv~K is an upper limit for the coronal or IGM pressure at a distance of 420~kpc. It follows from Equation \eqref{eq:hot} that the hot gas mass within $r=420$~kpc does not exceed $\sim 3 ֿ\times 10^{11}$~\msun, for a gas temperature $\gtrsim 2 \times 10^6$~K, comparable to the virial temperature of the Milky Way.

\subsubsection{Fitting the Half-Flux Radius}
\label{rhalf_fit}

\begin{figure}
\begin{tabular}{c}

 \includegraphics[width=0.45\textwidth]{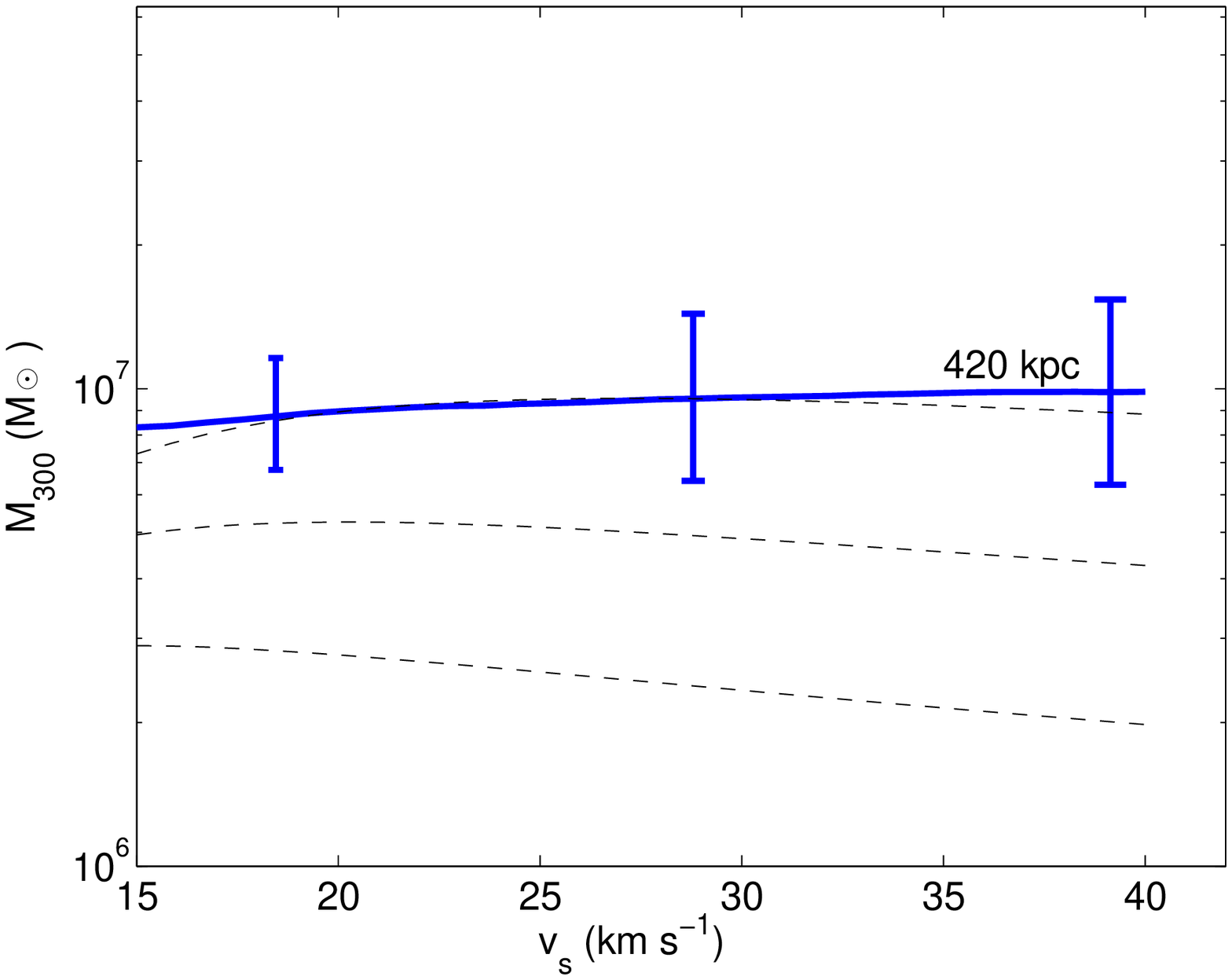} \\
 \includegraphics[width=0.45\textwidth]{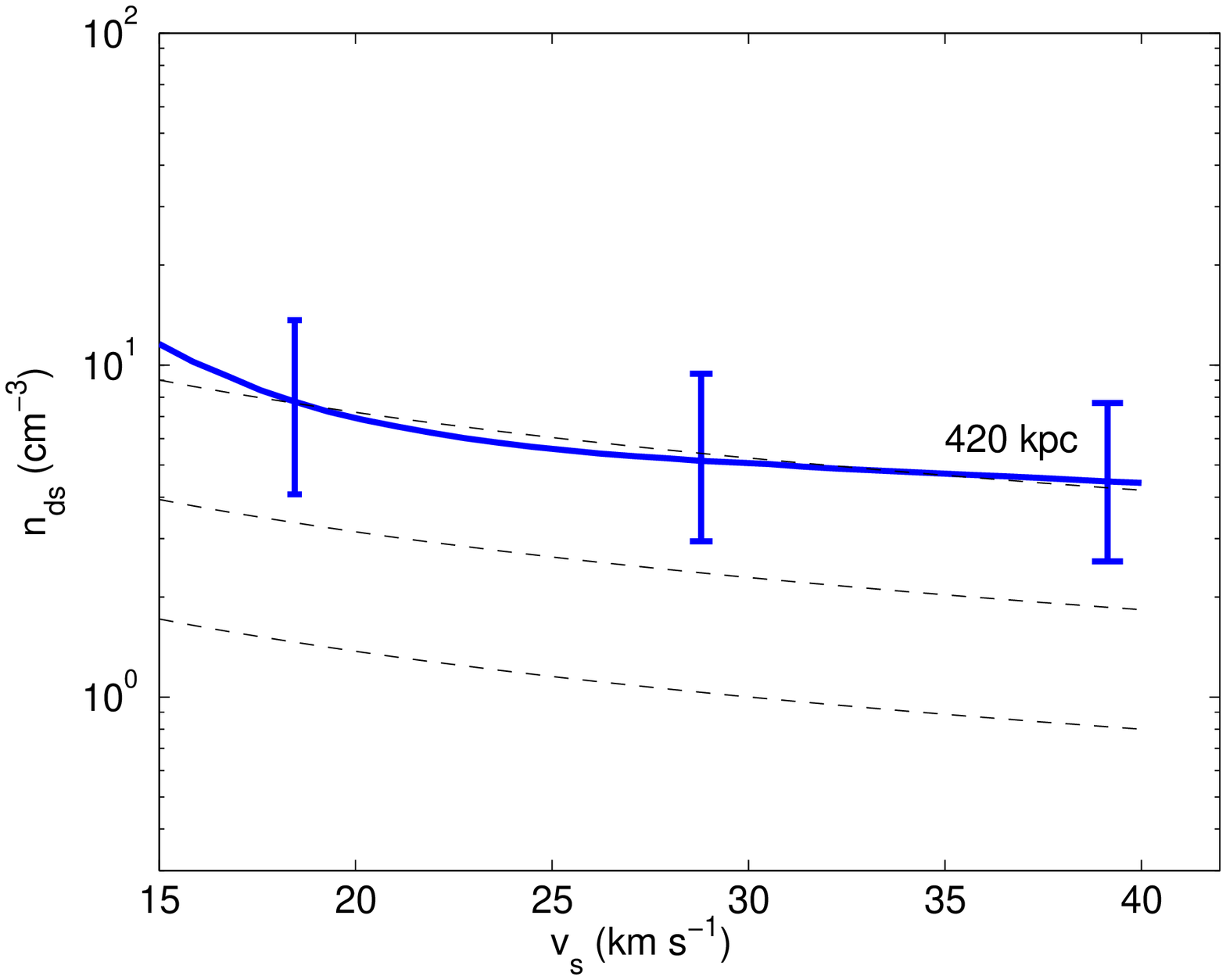} \\
\end{tabular}
\caption{Minihalo parameter combinations best reproducing the observed HI half-mass (projected) radius ($\rhalf  = 0.17$~kpc) of Leo~T, 
for \m300 versus \vs~(top) and \nds~versus \vs~(bottom) (for a distance of 420~kpc). 
The error bars show the range of models consistent with the observational $\Delta\rhalf$ (taken here to be 20\%). 
The dashed curves show \nds~and \m300~versus \vs~relations for \sig=-1, 0 and +1 ~halos.}
   \label{fig:leot_simple}
 \end{figure}

High resolution HI profiles for other systems, such as Leo~P or the UCHVCs, are not available.
For the near future, a more realistic goal is measurement of the half-flux 21~cm radii, as has been done for Leo~P. For optically thin (WNM) emission the half-flux radius is the projected HI half-mass radius, 
which can be either measured directly, or, for Leo~T, extracted from the full column density profile data. 
For given \vs~and \m300 we again set the total HI mass equal to the observed value
(as determined by the total 21~cm line flux) and then calculate the projected HI half-mass radius,
again assuming a gas temperature consistent with the line width (i.e., 6000~K for Leo~T).
We define the t-test parameter,
\begin{equation}
\eta = \frac{\vert r_{1/2,{\rm model}} - r_{1/2,{\rm obs}}\vert}{\Delta r_{1/2,{\rm obs}}}
\end{equation}
where $r_{1/2,{\rm obs}}$ and $r_{1/2,{\rm model}}$ are the observed and the calculated half-mass radii, 
and $\Delta r_{1/2,{\rm obs}}$ is the observational error. For Leo~T we find $r_{1/2,{\rm obs}} = 0.17$~kpc 
and estimate the error to be 20\%, typical for similar observations (Giovanelli, priv.~comm.).
We then minimize $\eta$ over the minihalo parameter space to find the best fitting models.  
Our fitting results for Leo~T are shown in Figure~\ref{fig:leot_simple}, for \m300 versus $\vs$ (top panel), and also for \nds~versus~\vs~(bottom).
The solid curves show the run of the best fitting models. The error bars show  the range in \m300 for which 
$\rhalf=r_{1/2,{\rm obs}} \pm \Delta r_{1/2,{\rm obs}}$, and hence $\eta = 1$. 
We find that the best fitting models are for $\m300 = 0.6-1.5 \times 10^7~\msun$, and that \vs~is not constrained.
The best fitting DM core densities are again 5 to 12 amu \cmv.
The consistency of these results with our fits for the full Leo~T profile supports our use of the total 21~cm flux 
and half-mass HI radius as probes of the halo potentials, as we do below for the UCHVCs.

We caution that the 21~cm observational sensitivity limit may affect the accuracy of the half-flux radius measurement. 
Examining our models, we find that detection of 90\% of the total 21~cm flux is enough to determine the half-flux 
radius with an error smaller than 10\% of the true value, calculated using the total flux. For our best models of Leo~T, 
the column density at the projected radius enclosing 90\% of the total HI mass is $\approx 3 \times 10^{19}$~\cmc. 
The radius at which the HI column density decreases to the detection limit of $2 \times 10^{19}$ encloses $\approx 95\%$ of the total HI mass. Thus, for Leo~T, the 21~cm sensitivity limit does not affect the \rhalf~estimate significantly.

\subsubsection{Limits on the Scale Velocity}
\label{vs_fit}

Although the halo scale velocity is not well constrained by the HI half-mass radius,
a lower limit can be set by the requirement that the WNM plus WIM gas mass be bound to the halo.
We apply two criteria for this. First, we require that the ratio of the potential to thermal energies per particle (SMW02~eq.~[34]), 
be larger than unity everywhere in the cloud, including in the outer WIM layers. 
Second, we require that the mean density,~$\bar{n}$, (including electrons and He) must be larger than the critical density, 
$n_{\rm crit} \equiv 2 \phim /\kb T$ where \phim~is the external bounding pressure.  This follows from 
the virial theorem for the gas, which may be written as $2(E_{\rm th}-E_{\rm surf})+W = 0$, where $E_{\rm th}$ is the thermal energy, 
$E_{\rm surf}$ is the surface pressure energy and $W$ is the gravitational energy. For a bound system, the total energy is negative, and
$E_{\rm tot} = E_{\rm th}+W<0$. Inserting $E_{\rm th} = \frac{3}{2}\kb TN$ and $E_{\rm surf} = \frac{3}{2}P_{\rm HIM}V$, 
we obtain our condition for a bound cloud
\begin{equation}  \label{eq:bound}
\bar{n} \equiv N/V > 2P_{\rm HIM}/\kb T \equiv n_{\rm crit} ~~~ .
\end{equation}  
Both of our criteria that the WNM+WIM gas be bound require $\vs \gtrsim 13 \kms$. 

Alternatively, the scale velocity for Leo~T may be estimated by a luminosity-to-mass abundance matching argument. 
For example, \citet{Kravtsov10} finds that for dwarf galaxies
\begin{equation}
L_{V} = 5 \times 10^3 \lsun \left( \frac{M_{\rm vir}}{10^9~\msun} \right)^{2.5} ~~~,
\end{equation}
where $M_{\rm vir}$ is the virial mass prior to accretion.
For Leo~T, $L_{V}=6 \times 10^4~\lsun$, giving $M_{\rm vir}=2.7 \times 10^9~\msun$. 
Given the \citet{Springel08} relation between \rmax~and \vmax~for unstripped subhalos
(see Equation [\ref{eq:rmax_corr2}]), and assuming that \vs~is not altered significantly 
during the accretion process, this virial mass corresponds to a median halo with \vs=33~\kms.

\subsection{UCHVCs}
\label{analysis_uchvc}

We now turn to the UCHVCs, which we wish to model as possible low-mass, gas-rich dwarf galaxies similar to Leo~T, at Local Group distances.
As we have just shown for Leo~T, the projected HI half-mass (i.e.~half-flux) radius may then be used to constrain the dark-matter halo parameters.
However, the UCHVC galaxy candidates are only barely (or not at all) resolved by the Arecibo beam.
We are thus motivated to construct models for the UCHVCs based on their total 21~cm fluxes as reported in the A13 UCHVC catalog,
to predict the projected half-mass radii and angular sizes for a range of assumed distances, as a guide to interpreting future high-resolution observations.

\subsubsection{Size-Distance Relations}

In calculating the HI half-mass radii for the UCHVCs we proceed as follows. 
For any observed total 21~cm flux, we convert to HI mass given an assumed distance $d$ 
\footnote{\mhi = $2.36 \times 10^5 \times S_{21} d_{\rm Mpc}^2~\msun$, where $S_{21}$ is the integrated 21~cm flux (Jy~\kms)}. 
We consider unstripped Burkert subhalo potentials with \sig~between -1 and 1, and \vs~from 20 to 40~km~s$^{-1}$. 
For each potential we find the total gas mass (including the outer ionized component) that reproduces the given HI mass, 
and we compute the hydrostatic gas density profile. Given the profile we extract the HI (projected) half-mass physical 
and angular radii as functions of the assumed distance.

We focus on the typical 21~cm fluxes for the 20 marginally-resolved and unresolved sources in the UCHVC catalog,
for which the median flux is 0.9 Jy~\kms, corresponding to an HI mass of $2.1 \times 10^5d^2_{\rm Mpc}$ \msun. 
We also compute models for 0.6 and 1.8 Jy~\kms, a range that covers 80\% of the $\sim 20$ most compact objects in the A13 catalog.

For the UCHVCs, with a typical 21~cm flux lower than in Leo~T, a higher sensitivity is needed to reliably measure the half-flux radii.
For our models we estimate that a column density detection limit of $\sim 10^{18}$~\cmc~is enough to enable recovery of $>90\%$ 
of the total HI mass for the range of distances we consider, thus reducing the systematic error in \rhalf~to $<10\%$.
For underconcentrated halos at distances smaller than 500~kpc, a higher sensitivity will be needed, of a few times $10^{17}$~\cmc. 

\begin{figure*}
 \includegraphics[width=0.95\textwidth]{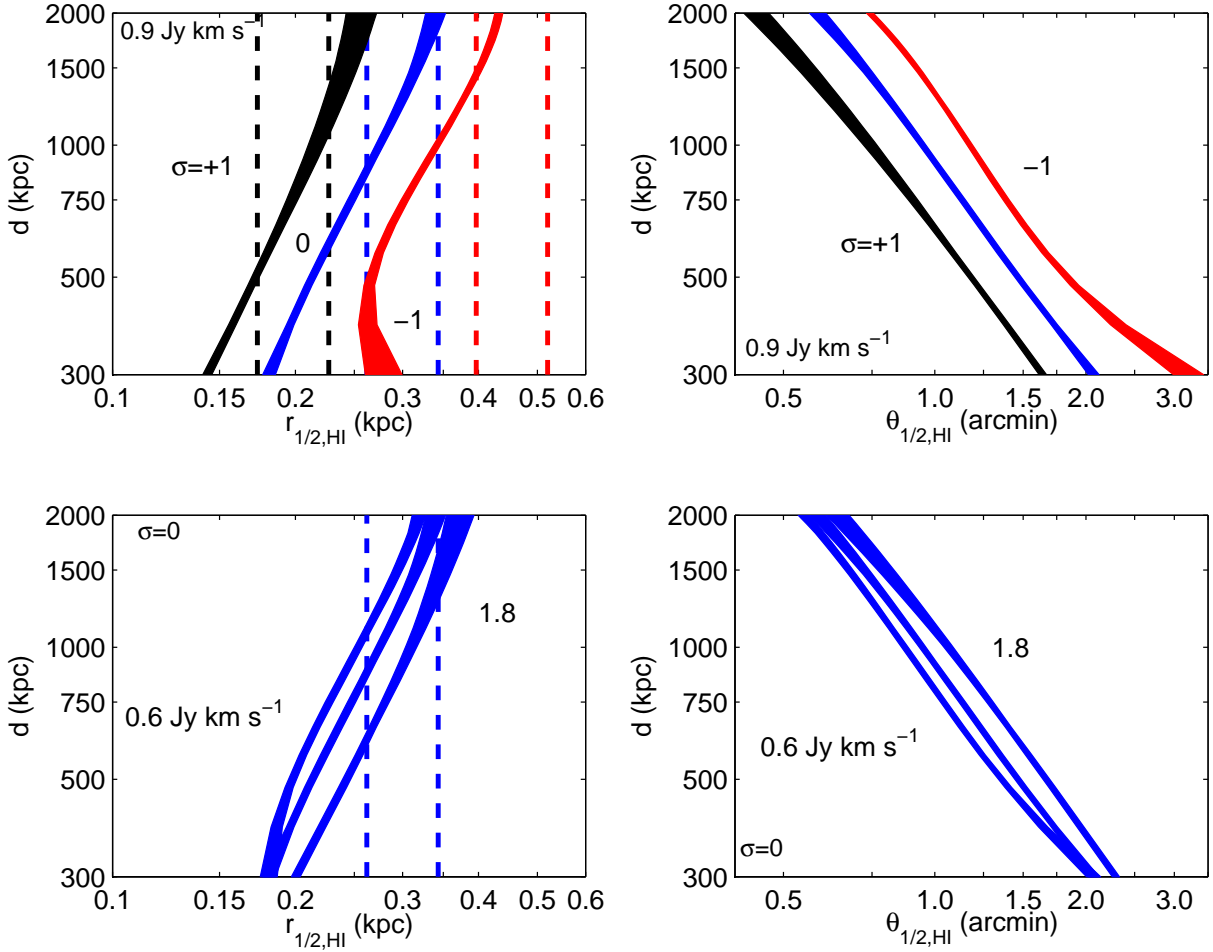} \\
\caption{Distance versus projected half-mass radius. Left panels are physical, right panels are angular sizes. 
Upper panels are for a (median) UCHVC flux of 0.9~Jy~\kms, for \sig=-1, 0 and +1 halos, with \vs~from 20 to 40~\kms. 
The dashed vertical lines show the range of analytic cloud sizes as given by Equation \eqref{eq:rgas2} for each \sig.
In the lower panels the 21~cm fluxes are 0.6, 0.9 and 1.8 Jy~\kms, and sizes are shown for \sig=0 halos.}
   \label{fig:rel}
 \end{figure*}

Our results for \rhalf($d$), and \thalf($d$) are displayed in Figure \ref{fig:rel}. We consider distances from 300~kpc to 2~Mpc. 
The resulting half-mass radii range from 0.15 to 0.45~kpc, corresponding to angular half-mass sizes from 0.4 to 3 arcminutes.
The upper two panels are results for the median UCHVC flux of 0.9~Jy~\kms, for \sig= -1, 0 and +1 halos.
For 0.9~Jy~\kms, the corresponding total HI gas mass ranges from $1.9 \times 10^4$ to $8.4 \times 10^5$~\msun~for 0.3 to 2~Mpc.
The lower two panels are for the three fluxes, 0.6, 0.9 and 1.8~Jy~\kms, for \sig=0 (median) halos.
For \sig=-1, 0 and +1, the characteristic dark-matter masses $\widetilde{M}_{300}^{\rm un}=2.9 \times 10^6$, $5.3 \times 10^6$, and $9.6 \times 10^6$~\msun.

In these computations departures from the small-$x$ limit and the photoionization cutoff 
play important roles in determining the cloud sizes.

In Figure~\ref{fig:rel} (left panels), the dashed lines show the range of physical cloud sizes as given by 
our small-$x$ analytic expression for \rhalf~(Equation~[\ref{eq:rgas2}]), for \vs~between 20 and 40~\kms~for each \sig,  and the (colored) strips show our numerically computed cloud sizes for this \vs~range. At large $d$ the HI gas masses and volumes are large and the half-mass radii extend
beyond the small-$x$ limits, especially for overconcentrated halos.  At small $d$ the clouds are within small-$x$ but the HI half-mass radii are reduced by photoionization, especially for underconcentrated halos. We find that for any distance, the cloud sizes depend mainly on the halo \sig. The dependence on 
\vs~is much weaker than indicated by our Equation~\eqref{eq:rgas2} because of departures from the small-$x$ limit for small \vs, for which the gas distribution becomes more extended than a simple Gaussian.

We can apply these calculations to Leo~P, with the reservation that our models do not include large scale rotation, which seems to be present, though not dominant, in this new member of the Local Group. We adopt the rough estimate of $1'$ from G13 as an approximate upper limit on the half-flux radius. For a median DM halo and the total 21~cm flux of 1.3~Jy~\kms, Equation \eqref{eq:dfit2} gives a lower limit on the distance, $d \gtrsim 1$~Mpc. Alternatively, for the optical distance estimate of 1.75~Mpc from R13, we can say that the 
DM halo hosting Leo~P has $\sig \gtrsim -1.6$. Modeling Leo~P more precisely should
incorporate the effects of rotation on the gas distribution.

\subsubsection{HI Gas Profiles}

\begin{figure*}
 \includegraphics[width=0.95\textwidth]{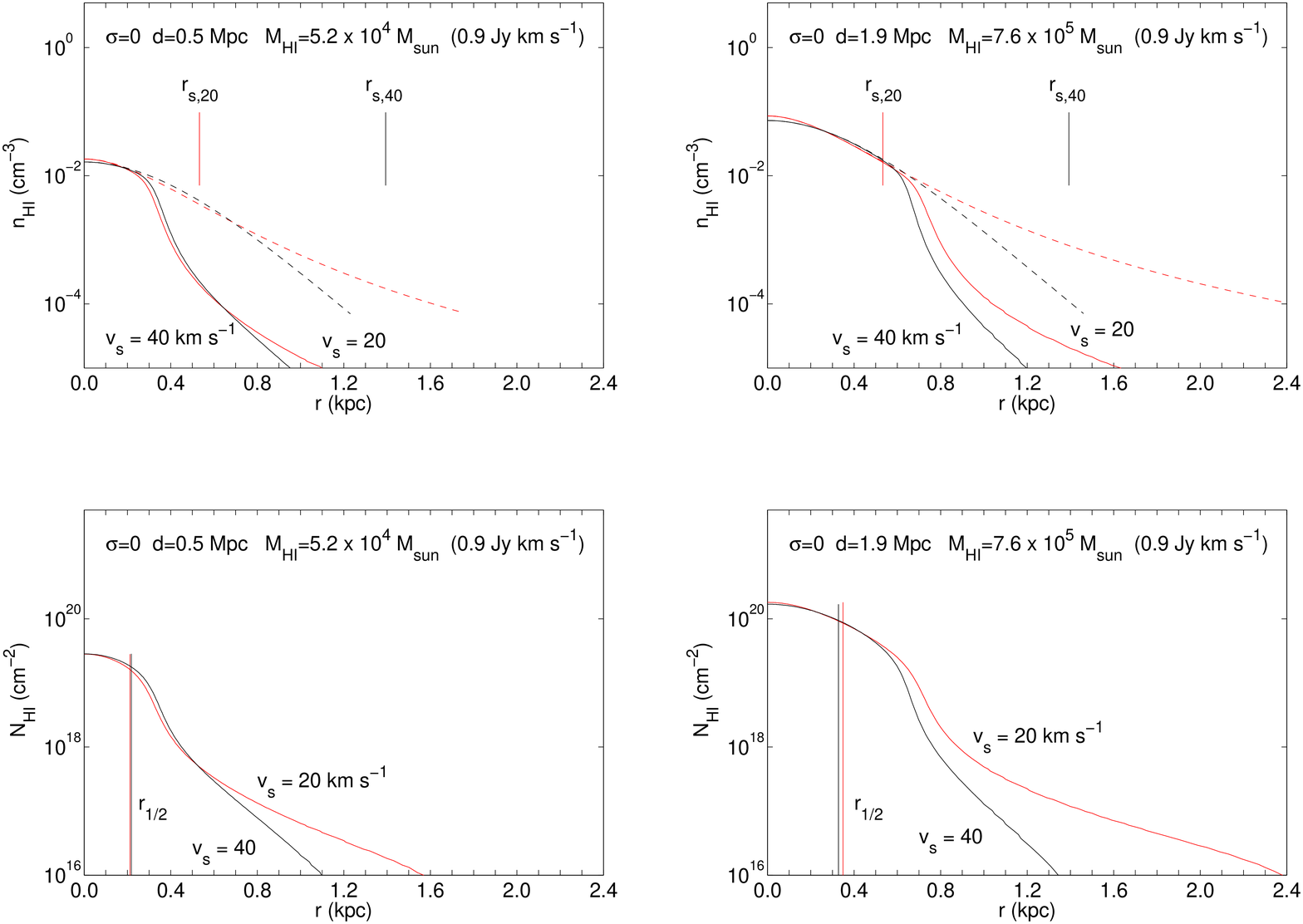}
\caption{HI gas density profiles for \sig=0 halos with \vs=20 and 40 \kms~(red and black curves)
for a UCHVC 21 cm flux of 0.9 Jy~\kms, and gas sound speed $c_g=8.2$~\kms.  
The halo scale radii are $\rs_{,20}=0.53$~kpc and $\rs_{,40}=1.39$~kpc.  
The upper panels show volume density profiles for the HI gas.  The lower panels show
the projected HI column densities.  Left panels are for $d=0.5$~Mpc
($\mhi=5.2\times 10^4$~\msun). Right panels are for $d=1.9$~Mpc ($\mhi=7.6\times 10^5$~\msun).
Solid curves are the numerically computed profiles including photoionization.
In the upper panels, the dashed curves are analytic hydrostatic gas distributions (footnote 5) for fully neutral gas. 
In the lower panels, the vertical lines show the computed half-mass radii of 0.22 and 0.35~kpc for $d=0.5$ and 1.9~Mpc.}
   \label{fig:prof1}
 \end{figure*}

\begin{figure*}
  \includegraphics[width=0.95\textwidth]{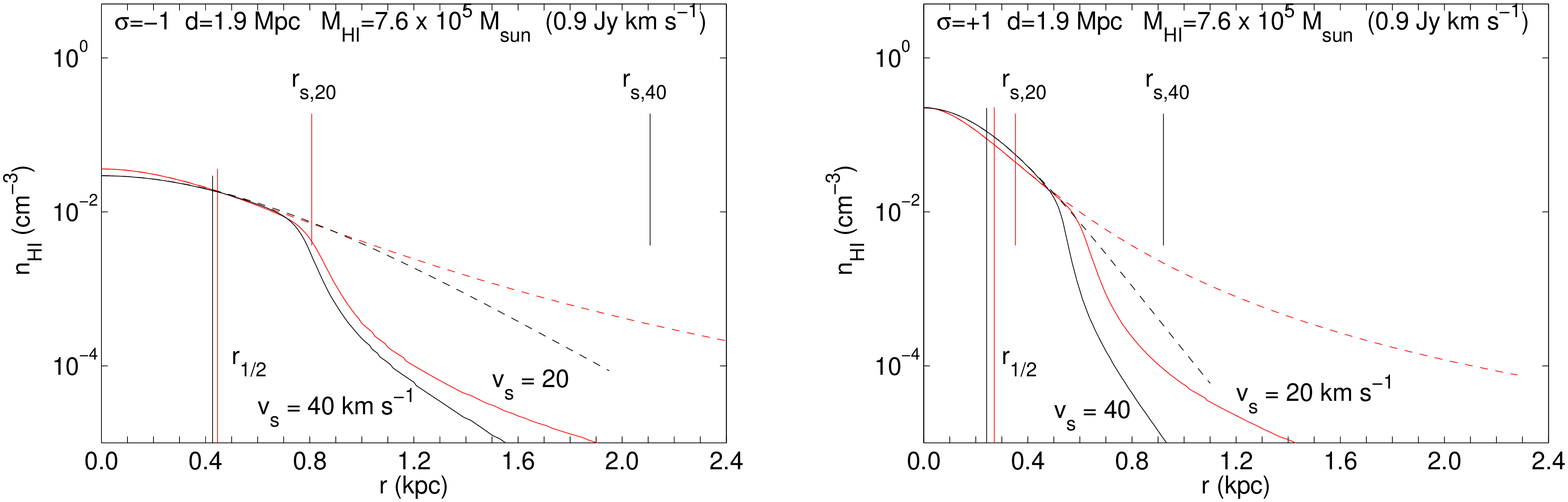}
 \caption{HI gas density profiles for $d=1.9$~Mpc and UCHVC 21 cm flux 0.9 Jy~\kms~($\mhi=7.6\times 10^5$~\msun).
 Left panel: Underconcenrated halos (\sig=-1) with $\vs=20$ and 40~\kms~and $r_{s,20}=0.81$~kpc and $r_{s,40}=2.11$~kpc.
 The HI half mass radius is 0.44~kpc. Right panel: Overconcenrated halos (\sig=+1) with $\vs=20$ and 40~\kms~and 
 $\rs=0.35$ and 0.92~kpc. The HI half mass radius is 0.25~kpc.}
   \label{fig:prof2}
 \end{figure*}

As shown in Figure \ref{fig:rel}, the size-distance relations are insensitive to \vs.
This behavior may be understood by considering Figures~\ref{fig:prof1} and \ref{fig:prof2}.
In Figure~\ref{fig:prof1} we display computed HI gas profiles (solid red and black curves)
for two \sig=0 halos with \vs =20~\kms~($r_{s,20}=0.53$~kpc, $n_{ds,20}=3.11$~amu, and \m300=5.3$ \times 10^6$~\msun)
and \vs =40~\kms~($r_{s,40}=1.39$~kpc, $n_{ds,40}=1.81$~amu, and \m300=$4.3 \times 10^6$~\msun).
We show HI profiles for two distances, $d=0.5$ and 1.9~Mpc (left and right panels),
for which the HI masses are $5.2\times 10^4$ and $7.6\times 10^5$~\msun, for a 21 cm flux of 0.9 Jy~\kms.  
The upper panels in Figure~\ref{fig:prof1} show the HI gas volume density (\cmv) and the lower panels show the 
projected HI column density (\cmc), as functions of radial offset $r$ (kpc) from the minihalo centers.  The sharp drops in the HI density profiles indicate
the locations of the ionization fronts.  The HI gas profile shapes as determined by the DM halo potential, 
and the given (observed) HI masses for each distance uniquely determine the central gas densities for each model.  

The dashed curves in the upper panels of Figure~\ref{fig:prof1} show the gas-density profiles
as given by the general SMW02 expression for hydrostatic distributions in Burkert 
halos~\footnote{\ For gas in Burkert halos and for all $x\equiv r/r_s$ the hydrostatic gas distribution is
$f_{\rm gas}(x) =\bigl[{e}^{{-(1+1/x){\rm tan}^{-1}x}}(1+x)^{(1+1/x)}(1+x^2)^{(1/2)(1/x-1)}\bigr]^{{{(3/2}(v_s/c_g)^2}}$
(SMW02, Table 5). For $x \ll 1$, $f_{\rm gas}(x)$ approaches the Gaussian form $f_{gas} = e^{-\frac{1}{2}(\vs/\cg)^2 x^2}$ (Equation~[\ref{eq:rgas1}]).}
for constant gas velocity dispersion \cg, here set to 8.2~\kms~for fully neutral gas.
The red and black dashed curves are for $\vs=20$ and 40~\kms~respectively.  
For $d=1.9$~Mpc, it is apparent that the neutral core for $\vs=20$~\kms~extends beyond the small-$x$ limit ($r/\rs \ll 1$). 
For $\vs=20$~\kms, the ionization front is at 0.73~\kpc, larger than $r_{s,20}=0.53$~kpc. 
However, for $\vs=40$~\kms~the neutral gas cloud is limited to small-$x$ and the IF is at 0.67~\kpc, smaller than $r_{s,40}=1.39$~\kpc.
Our numerically computed HI gas distributions (solid curves) match the analytic curves (dashed) out to the locations of the ionization fronts. 
At larger radii the HI gas profiles drop below the dashed curves because of the reduced neutral fractions.

The analytic gas distribution (footnote 5), and the numerically computed HI gas distribution within the neutral cores,
have the remarkable property that $f_{\rm gas}(x,\vs=20)$ is almost  identical to $f_{\rm gas}(x/2.6,\vs=40)$ up to $x\approx 1.5$. 
Here $r_{s,40}/r_{s,20}=2.6$, so that, $f_{\rm gas}(r,\vs=20)\approx f_{\rm gas}(r,\vs=40)$ 
up to $r\approx 0.8$~kpc, as indeed shown by the curves. At larger radii the dashed curves diverge. A radius $r=0.8$~kpc is sufficiently large (for \sig=0)
to contain the neutral cores for all distances between 300~kpc and 2~Mpc, and the neutral gas profiles are therefore essentially independent of \vs. 
As \vs~is reduced from 40 to 20~km~s$^{-1}$ the gas becomes more extended than a simple small-$x$ Gaussian. This compensates for the reduction in the 
expected small-$x$-Gaussian sizes that are proportional to \rs/\vs. Thus, for a given distance, the only way to alter the cloud sizes significantly is 
to change the scale radius \rs~by varying \sig~rather than by varying \vs. This is illustrated in Figure~\ref{fig:prof2} in which we plot
$\vs=20$ and 40~km~s$^{-1}$ profiles for \sig=-1 and +1 halos (for $d=1.9$~Mpc).
For $\sig=-1$, $\rs = 0.81$ and 2.11~kpc, $\nds=1.35$ and 0.79~amu and $\m300=2.77 \times 10^6$ and $1.99 \times 10^6$~\msun, for $\vs=20$ and 40~\kms.
For $\sig=+1$, $\rs = 0.35$ and 0.92~kpc, $\nds=7.11$ and 4.14~amu and $\m300=8.92 \times 10^6$ and $8.87 \times 10^6$~\msun, for $\vs=20$ and 40~\kms.
The gas clouds in the overconcentrated halo are smaller, and in the underconcentrated halos they are larger, compared to the clouds in median halos.  
The sizes are again very weakly dependent on \vs. For $d=1.9$~Mpc, $r_{1/2}\approx 0.44$, 0.34, and 0.25~\kpc, for \sig=-1, 0, and +1.

Figure~\ref{fig:prof1} shows that for our distance range the central gas densities are high enough for the formation of neutral cores 
\footnote{For small-$x$ Gaussian gas distributions the critical central gas densities for the formation of neutral cores are given 
by the emission-measure condition $r_{\rm gas} n^2_{-3}= 0.99 J^*_3$ where $J_3^*$ is the ionizing photon
intensity in units of 10$^3$~photons~cm$^{-2}$~s$^{-1}$~sr$^{-1}$ and $r_{\rm gas}$ in kpc
(SMW02 Equation~[A4]). For $r_{\rm gas}\equiv \sqrt{2}c_g\rs/\vs\approx 0.2$~kpc, with \cg=11.6 \kms~for WIM, 
and $J^*_3=1.02$ for the metagalactic radiation, neutral cores appear for central gas densities $\gtrsim 2\times 10^{-3}$~cm$^{-3}$.}.
For lower distances (smaller masses) the central gas densities are lower, and the ionization fronts occur at smaller radii. 
The half-mass radii are therefore also reduced. For $d=1.9$ to 0.5~Mpc, \rhalf~decreases from 0.35 to 0.22~\kpc. 
For higher 21~cm fluxes and cloud masses the neutral core truncation points occur at larger radii for any distance, and the cloud sizes are increased
as shown in the lower panels of Figure~\ref{fig:rel}. For any flux, the size versus distance relation depends mainly on the halo \sig.

These numerical results can be described by simple power law relations. For varying sigma (upper right hand panel in Figure~\ref{fig:rel}), we can write
\begin{equation} \label{eq:dfit1}
	d = 0.91 \times 10^{-0.15\sig} \times \left( \frac{\thalf}{1'} \right)^{-1.47} ~~~ \rm Mpc,
\end{equation}
and for varying flux for \sig=0 subhalos (lower right hand panel),
\begin{equation} \label{eq:dfit2}
		d = 0.91 \times \left( \frac{S_{21}}{0.9 {\rm~ Jy} \kms} \right) ^{0.24}\times \left( \frac{\thalf}{1'} \right)^{-1.47} ~~~ \rm Mpc.	
\end{equation}
Both expressions are accurate to within 10\% for \sig~between -1 and +1, fluxes from 0.6 to 1.8~Jy~\kms , 
and for angular half-flux radii between $0.6'$ and $2.5'$. In these relations we assume that the 
observational sensitivity limit is sufficient for an accurate determination of \thalf, as discussed above.

Figure~\ref{fig:rel} and Equations \eqref{eq:dfit1} and \eqref{eq:dfit2} may be used to estimate distances to the UCHVCs as minihalos given the observed 21~cm fluxes and angular sizes.  
For example, Figure~\ref{fig:rel} shows that a UCHVC embedded in a median unstripped subhalo, with an HI size $\thalf=1$' and 21~cm flux of $\sim 0.9$~Jy~\kms, 
has $\rhalf~ \sim 0.26$~kpc, and is hence at a distance of $\sim 0.91$~Mpc. 
Assuming underconcentrated halos, with \sig=-1, will increase the size and distance to $\sim 0.37$~kpc and $\sim 1.28$~Mpc, respectively. 
Switching to overconcentrated halos, with \sig=+1, will reduce the physical size to $\sim 0.18$~kpc and place the object at $\sim 0.63$~Mpc. 
For comparison, decreasing the 21~cm flux by 30\%, to 0.6~Jy~\kms~will reduce the physical size of a median halo to $\sim 0.23$~kpc and place it at $\sim 0.83$~Mpc. 
Doubling the flux, to 1.8~Jy~\kms, will change the size and distance to $\sim 0.31$~kpc and $\sim 1.08$~Mpc, respectively. 
Alternatively, if the distance to the UCHVCs is measured by any other method (as for Leo~P), the physical scale of gas distribution can be 
used to constrain the DM halo properties in which the UCHVCs may reside, as we have done for Leo~T.

\section{Summary \& Discussion}
\label{sec_summary}

In this paper we have presented dark-matter minihalo models for the HI gas in the 
ultra-compact high-velocity clouds (UCHVCs) detected as part of the 21~cm ALFALFA survey 
\citep{Gio10,Adams13}, and for the resolved HI gas distribution \citep{RW08} in the Local Group dwarf galaxy Leo~T.

Our minihalo cloud models are based on those presented in \citet[SMW02]{SMW02}. We consider thermally supported, hydrostatic, 
non-rotating, $\sim 10^4$~K HI clouds, embedded in gravitationally dominant DM potential wells, with negligible external pressure.  
The observable (21~cm) WNM cores are surrounded by WIM shielding envelopes, also bound to the minihalos. 
We assume that the WIM is photoionized by the present-day ($z$=0) UV/X-ray metagalactic radiation field.   

In \S~\ref{sec_conf} we discuss pressure versus gravitational confinement for the UCHVCs as Local Group objects,
and show how the observed HI column densities and angular sizes may be used to set limits on the ambient hot gas pressures and masses.

In \S~\ref{sec_models_dm} we assume (observationally-based) ``flat-core" \citep{Burk95} DM density profiles,
and adopt the cosmological correlation between halo structural parameters appropriate for 
subhalos within Galaxy-scale parent halos at redshift $z=0$, as found in simulations \citep{Springel08, Rocha13}.
In \S~\ref{sec_mass_scale} we show that typical (\sig=0) tidally stripped minihalos always contain a DM mass of  
$\sim 0.9 \times 10^7$~\msun~within the central 300~pc, weakly dependent on subhalo scale-velocity (or maximal circular velocity), and total halo mass.
Our flat-core minihalos (stripped) thus naturally reproduce the Strigari mass of $\m300 \sim 10^7$~\msun~found 
via optical stellar velocity dispersion measurements in low-mass Local Group dwarfs (S08). 
We also show that the mass-discrepancy between simulated and observed dwarf-galaxies 
noted by BK11 for NFW halos is resolved if the flat-core profile is adopted.

In \S~\ref{sec_models_gas} we present analytic expressions relating the gravitationally-confined HI cloud properties to the minihalo parameters, including the dependence of the projected HI half-mass radii on the minihalo \sig~and scale-velocity \vs~(Equation~[\ref{eq:rgas1}]). These expressions are derived in the ``small-$x$ limit" where all of the neutral gas is contained well within the halo scale radius, and can be used for rough estimation of object sizes and distances.

In \S~\ref{sec_analysis} we present numerical model computations for the gas distributions in Leo~T and the UCHVCs 
without {\it a priori} assumptions on the extents of the clouds, and including the truncation of the neutral cores by external photoionization by metagalactic radiation. 
We start with Leo~T (\S~\ref{analysis_leot}) and fit the entire observed HI column density profile (Figure~\ref{fig:leot_prof}),  
assuming a WNM temperature of 6000~K as determined from the observed 21~cm line width, and given the total observed HI mass of $2.8\times 10^5$~\msun.  
We assume the external bounding pressure is negligible compared to the gravitational potential.
The observed HI profile provides an upper limit on the pressure, above which the outer HIM will penetrate into the halo and alter the profile shape. 
For Leo~T the limit is $\phim \lesssim 150$~\cmv~K, and this is then a constraint on the IGM thermal pressure in the Local Group at distances $420$~kpc.
For $T_{\rm HIM} \gtrsim 2\times 10^6$~K, comparable to the virial temperature of the Milky Way, the mass of hot gas within 420~kpc is $M_{\rm HIM} \lesssim 3 \times 10^{11}$~\msun.
Our models also predict the presence of CNM in the Leo~T core, consistent with observations. 
The observed HI profile sets a strong constraint on \sig~and \m300.
The best fitting profiles are for halos with $\m300=8.0~(\pm 0.2) \times 10^6$~\msun, only weakly dependent on scale-velocity \vs.
For unstripped halos this mass corresponds to $\sig \approx +0.75$.
We recover consistent \m300~and \sig~if we solve for just the HI half-mass radius (which for Leo~T is 0.17~kpc).  
In this procedure we again set the total HI mass and temperature to the observed values, and adjust the halo parameters to match the observed half-mass radius. 
Our detailed fits for Leo~T show that the projected HI half-mass radius may be used reliably to constrain the halo parameters.

The recent discovery that the UCHVC HI102145.0/Leo~P is a dwarf galaxy (\citealt{Gio13,Rhode13}), 
in addition to our minihalo fits for the HI gas profile in Leo~T, suggest that at least some of the UCHVCs discovered in the ALFALFA survey may in fact be gas rich but optically very faint dwarf galaxies, also in the Local Group. 
The orginal SMW02 study focused on the much larger CHVCs (diameters $\gtrsim 24'$).
Modeling the CHVCs as median-stripped-minihalos, the implied distances are $< 200$~kpc (as concluded by SMW02)
but it now seems less likely that at such distances they could retain their gas, as indicated by the nearby gasless Sloan dwarfs. 
The UCHVCs are smaller and hence could be much more distant. Thus they are more plausible dwarf galaxy candidates.

For the UCHVCs (\S~\ref{analysis_uchvc}) we calculate and plot (Figure \ref{fig:rel}) predicted half-mass HI radii (physical and angular) for assumed distances between 0.3 and 2 Mpc, for 21~cm fluxes between 0.6 to 1.8~Jy~\kms, as observed in the more compact objects in the A13 UCHVC catalog. We consider unstripped subhalos within $\sig=\pm 1$ of the cosmological median, with scale velocities \vs~from 20 to 40 \kms. We find that the cloud sizes depend mainly on $\sigma$ and only weakly on \vs. For a typical UCHVC with a 21~cm flux equal to 0.9~Jy~\kms~embedded in a median (\sig=0) unstripped subhalo, we predict physical HI half-mass radii of 0.18 to 0.35 kpc (or angular sizes of 0.6 to 2.1 arcmin) for this distance range. If the angular half-flux radii of the UCHVCs are measured,  our predictions can be used to estimate the (currently unknown) distances to these objects. 
Alternatively, if the distances to the UCHVCs are measured by some other method, future high-resolution observations of the HI sizes will be useful in constraining the halo parameters, as we have done for Leo~T.

The {\it Via~Lactea~II} simulations (\citealt{Diemand07}) indicate that there are $\sim 170$ halos with  \vs~between 15 and 50~\kms~at distances between 300~kpc and 1~Mpc from a MW-sized galaxy. This suggests that if only a fraction (10-20\%) of the UCHVCs are indeed dwarf galaxies, and assuming that the ALFALFA survey-area is representative of the rest of the sky,  these objects could resolve the missing satellite problem for this mass range.

To summarize our conclusions and predictions: First, we predict that searches for stars in the most compact UCHVCs
in the A13 catalog and in other surveys will show that some of these objects are optically faint, gas-rich dwarf galaxies.  
Second, high-resolution (sub-arcmin) HI mapping observations of such gas-rich UCHVC/dwarfs 
will show that they obey the size-versus-distance relations as given by our Equations \eqref{eq:dfit1} and \eqref{eq:dfit2}. 
For a Local Group distance of 1~Mpc, our predicted characteristic projected HI half-mass radius is $1'$ 
for $\sim 10^4$~K WNM in a median unstripped subhalo obeying the \rmax~versus \vmax~relations found in simulations. 
Third, based on our analysis of the resolved HI gas distribution in Leo~T we conclude that at 
a distance of $\sim 400$~kpc from the Galaxy the thermal pressure of any ambient hot coronal
or IGM gas is less than $150$~\cmv~K, implying a hot gas mass of less than $3\times 10^{11}$~\msun~within this radius. 
Fourth, we predict that at distances beyond the virial radii (unstripped) dwarf galaxies
around the Milky Way and Andromeda will have characteristic inner masses \m300~of $\sim 5\times 10^6 $~\msun~(but with scatter), 
and lower by a factor $\sim 2$~compared to the nearby (stripped) Sloan dwarfs.
Fifth, we predict that as ever lower-luminosity dwarf galaxies are observed within the virial radius, 
\m300~will remain invariant, since in our picture the characteristic mass does not reflect a star-formation threshold, 
but rather the independence of \m300~on subhalo mass for flat-core DM profiles,
from scale velocities of 50~\kms~down to $\sim 15$~\kms.

We have focused on flat-cored subhalos in this paper because 
(a) they are observed in at least some low-mass galaxies,
(b) they naturally give rise to the observed common mass scale, and
(c) because flat-cores resolve the mass-discrepancy problem found in DM-only simulations of Local Group structure growth.
\citet{Zolotov12} show that stellar feedback processes combined with tidal stripping 
can flatten cores with initially diverging NFW density distributions. 
Alternatively, \citet{Rocha13} and \citet{Zavala13} suggest that self-interacting-dark-matter (SIDM) may also produce flat-core 
DM profiles on small scales but leave the halos unaffected on larger scales. 
We suggest that optically-faint but gas-rich dwarf galaxies - perhaps UCHVCs as dwarf galaxies - 
may be the best objects to study to distinguish between these theoretical options, since feedback effects
in these almost starless systems are expected to be of minor importance to the evolution of their DM halos (see \citealt{Pen12}). 
Hence, existence of flat DM cores in UCHVCs (or in Leo~T and Leo~P), if proved, may be intrinsic and evidence for SIDM.
Sensitive high resolution 21~cm observations of the most compact UCHVCs, as well as searches for stars and stellar features in these objects, would be worthwhile.

\begin{acknowledgements}

We thank Leo Blitz, James Bullock, Sukanya Chakrabarti, Orly Gnat, Andy Gould, Avi Loeb, Chung-Pei Ma, David Spergel and Benny Trakhtenbrot for helpful discussions, and Elizabeth Adams, Riccardo Giovanelli, and Martha Haynes for providing us with their UCHVC data in advance of publication and for their helpful comments on this paper. We thank Emma Ryan-Weber for sending us her Leo~T HI profile in digital form. The referee was very helpful and provided constructive comments that improved our paper. We thank the DFG for support via German-Israeli Project Cooperation grant STE1869/1-1/GE625/15-1. Our research is supported by a PBC Israel Science Foundation I-CORE Program 1829/12. C. F. M. is supported in part by NSF grants AST-0908533 and AST-1211729.

\end{acknowledgements}

\bibliographystyle{plainnat}

\end{document}